\documentclass[12pt]{iopart}
\usepackage{amsmath}
\usepackage{graphicx}
\usepackage{float}
\usepackage{algorithm}
\usepackage{algpseudocode}
\usepackage{pdflscape}
\usepackage{lipsum}
\usepackage{booktabs} 
\usepackage{xcolor}
\usepackage{pifont}
\usepackage{array}
\usepackage{url}

\newcolumntype{L}[1]{>{\raggedright\let\newline\\\arraybackslash\hspace{0pt}}m{#1}}
\newcolumntype{C}[1]{>{\centering\let\newline\\\arraybackslash\hspace{0pt}}m{#1}}
\newcolumntype{R}[1]{>{\raggedleft\let\newline\\\arraybackslash\hspace{0pt}}m{#1}}

\begin{document}

\title{Analyzing Warp Drive Spacetimes with Warp Factory}

\author{Christopher Helmerich${}^{1,2}$, Jared Fuchs${}^{1,2}$, Alexey Bobrick${}^{2,3}$, \ \ \ Luke Sellers${}^{2,4}$, Brandon Melcher${}^{2}$, \& Gianni Martire${}^{2}$}

\address{{${}^{1}$The University of Alabama in Huntsville, 301 Sparkman Drive,
Huntsville, Alabama, 35899, U.S.}}
\address{${}^{2}$Advanced Propulsion Laboratory at Applied Physics, 477 Madison Avenue, New York, 10022, U.S.}
\address{${}^{3}$Technion - Israel Institute of Technology, Physics Department, Haifa 32000, Israel}
\address{${}^{4}$UCLA Department of Physics \& Astronomy,
475 Portola Plaza, Los Angeles, CA 90095, U.S.}

\ead{cdh0028@uah.edu, christopher@appliedphysics.org, and christopher.d.helmerich@gmail.com}

\vspace{10pt}

\begin{abstract}

The field of warp research has been dominated by analytical methods to investigate potential solutions. However, these approaches often favor simple metric forms that facilitate analysis but ultimately limit the range of exploration of novel solutions. So far the proposed solutions have been unphysical, requiring energy condition violations and large energy requirements. To overcome the analytical limitations in warp research, we introduce Warp Factory: a numerical toolkit designed for modeling warp drive spacetimes. By leveraging numerical analysis, Warp Factory enables the examination of general warp drive geometries by evaluating the Einstein field equations and computing energy conditions. Furthermore, this comprehensive toolkit provides the determination of metric scalars and insightful visualizations in both 2D and 3D, offering a deeper understanding of metrics and their corresponding stress-energy tensors. The paper delves into the methodology employed by Warp Factory in evaluating the physicality of warp drive spacetimes and highlights its application in assessing commonly modeled warp drive metrics. By leveraging the capabilities of Warp Factory, we aim to further warp drive research and hopefully bring us closer to realizing physically achievable warp drives.

\end{abstract}

%
\vspace{2pc}

\submitto{\CQG}

\clearpage
%
%
%

\section{Introduction}

Warp drive research has been mostly driven by analytic studies of various metrics. Such limitations have led to a narrow perspective when it comes to warp drive physicality, especially due to the non-linear nature of the gravitational field equations. We developed Warp Factory to broaden the view of warp drive research. Our numerical toolkit can evaluate the Einstein equations to facilitate the analysis of the stress-energy tensor and its consequent energy condition satisfaction or violation.

The first warp drive solutions developed by Alcubierre \cite{1994CQGra..11L..73A} and then further developed by Van Den Broeck \cite{1999CQGra..16.3973V} enabled passengers inside the warp bubble to move at arbitrary speeds but at the cost of both enormous energy density and unphysical stress-energy tensors using negative energy. The recent solutions from Lentz \cite{2020arXiv200607125L} and Fell \cite{2021CQGra..38o5020F} both find solutions with positive Eulerian energy density. However, it has been argued that these still both violate the weak energy condition, as shown generally by \cite{2022PhRvD.105f4038S} and by Fell himself. Since the publication of \cite{2021CQGra..38j5009B} an extended class of possible solutions has been proposed for warp drives, with an emphasis on physical solutions which require subluminal speeds. 
 
The key challenge in warp research is two-fold: The first is evaluating the complicated Einstein field equations to find stress-energy tensors and the second is in evaluating those stress-energy tensors for physicality. Proving physicality is a unique focus in the field of warp drive research that is often not required when solving general astrophysical problems. For simple warp drives, such as those with symmetry and time-invariance, analytical solutions for the stress-energy tensor can usually be easily determined. Once the analytic stress-energy tensor is known, the physicality of the solution is then evaluated by contracting the stress-energy tensor with the velocity of different observers. Often the contraction process to show physicality is incorrectly limited to a single, easily-defined observer. For simple metrics, this is typically sufficient to prove that a solution is not physical. In a general case, physicality is more difficult to prove and single observers are not sufficient. The broader cases of warp solutions, such as those with Schwarzschild boundaries, asymmetries, and accelerations, will have complicated metrics that are inaccessible to analytical approaches. This presents a significant challenge to the development of new warp solutions.
 
To overcome the difficulties of creating and analyzing general warp drives, Warp Factory\footnote{\url{https://github.com/NerdsWithAttitudes/WarpFactory}.}, a numerical analysis toolkit, has been developed that provides a general numerical evaluation of the Einstein field equations and physicality of the stress-energy tensor across a range of observers for arbitrary input metrics. Along with testing for physicality, the toolkit can provide other analyses, such as constructing metric scalars and analyzing momentum flows. This numerical approach offers robust methods to evaluate physicality and provides the ability to visualize the metric and stress-energy tensors in order to gain valuable insights.

In this paper, we establish the methods used in Warp Factory to construct and analyze spacetimes and demonstrate this analysis on several example warp drive metrics. Warp theory and the definition of physicality are introduced in Section \ref{sec:warptheory}. The evaluation of the Einstein field equations, the methods used in conducting stress-energy tensor analysis, and the solution of metric scalars are explained in Section \ref{sec:methods}. Then, in Section \ref{sec:metricEval}, these methods are applied to evaluate metrics commonly considered in warp literature. Finally, Section \ref{sec:discussion} summarizes and discusses the results, and Section \ref{sec:conclusion} concludes with closing remarks.

\subsection{Key Results}
\begin{enumerate}
    \item Numerical methods are established and demonstrated for evaluating the physicality of metrics using a custom framework in MATLAB called Warp Factory. 
    \item The Alcubierre, Van Den Broeck, Bobrick-Martire Modified Time, and Lentz-inspired warp solutions are evaluated, showcasing their stress-energy components, energy condition violations, and metric scalars.
\end{enumerate}

\section{Warp Theory}\label{sec:warptheory}
The field of warp mechanics does not have a unified definition of what is required by a warp solution. However, there are several aspects that emerge as important features of any solution. A warp drive aims to provide a controlled and comfortable journey through space. This goal motivates the three core features of a warp drive: geodesic transport, a flat passenger region, and a spatially compact and comoving warp bubble.
\begin{enumerate}
\item \textit{Geodesic transport.} Warp drives should transport passengers from point A to point B along a geodesic trajectory. This means the passengers inside the warp drive do not experience local acceleration while being transported\footnote{In the case that a local acceleration is desired, for example, 1g, the statement becomes the local acceleration of passengers should be limited.}. For a non-trivial solution, the passengers should not `already be going to' point B. One possible scenario is that passengers start at rest (relative to points A and B) at point A, are transported to point B, and are then, again, at rest relative to both points. Additionally, this transport should occur for a limited region in spacetime.
\item \textit{Empty passenger region.} There should exist a defined passenger volume that is both vacuum\footnote{Ignoring the mass of the passengers.} ($T^{\mu\nu} = 0$) and free from large tidal forces.
\item \textit{A spatially bounded, comoving bubble.} The stress-energy required for the geodesic transport should not extend to infinity\footnote{Perhaps some energy could be radiated to infinity but that energy should be causally connected to the bubble.}. The stress-energy of the bubble also needs to move along with the transported observers. The requirement of it comoving with passengers distinguishes warp drive solutions from Krasnikov tubes \cite{1997PhRvD..56.2100E}.
\end{enumerate}

This framework establishes the primary goals of any generic warp solution. The warp metrics proposed in the literature have all satisfied different features in this list. For example, Alcubierre has a flat passenger volume and a warp bubble whose spatial extent is determined by a shape parameter\footnote{While the functions used by Alcubierre are not exactly zero at $r \gg R_{bubble}$, it is, for practical purposes, approximately compact. This is also true for the passenger volume flatness.}.

\subsection{Conception of Warp}\label{sec:warpconcept}
In Star Trek and other science fiction, the notion of a warp drive has several distinct features which may or may not apply to real, physical, warp drives including faster-than-light movement, self-acceleration, a normal rate of time-passage, and possessing a bubble of energy that surrounds a ship.

\begin{itemize}

\item \textit{Moving Faster-than-light:}
The creation of warp has historically been motivated by the ability to potentially travel at superluminal speeds, but the broader solutions of warp drives are not restricted to this criteria and interesting solutions for subluminal speeds also exist. In the recent paradigm by Bobrick \cite{2021CQGra..38j5009B}, warp drives are classified into regimes of solutions based on speed. Likely, physical solutions are restricted to the subluminal regime \cite{2021CQGra..38j5009B}.

\item \textit{Self-acceleration:}
To transport observers at rest to some other point in spacetime requires a coordinate acceleration as measured by an external observer. For objects in flat space, acceleration requires the transfer of momentum of the accelerating system. Within a curved spacetime, observers can travel along geodesics which can be perceived by other observers as a relative acceleration. In addition, the constraint of a comoving bubble also requires the bubble itself to self-accelerate or be conventionally propelled.

\item \textit{Passenger Time:}
In line with the Star Trek conception of warp drives, the first warp solution from Alcubierre was constructed such that the passengers and external observers both experience the same passage of time (proper time). This construction exists in most of the solutions developed. However, allowing the modification of the lapse rate might be an important part of physical solutions. Warp shells with a bubble consisting of regular matter, such as a matter shell, may require a change in lapse to the passenger volume \cite{PhysicalWarp2023}.

\item \textit{Spaceships and Warp Bubbles:}
In Star Trek, the ship creates some kind of radiated energy field that dissipates when the warp effect is removed. This does not need to be the case in general. For example, a bubble might be constructed by a shell of matter whose stress-energy tensor changes during the warp journey but returns back to a regular matter shell when the warp effect is removed.

\end{itemize}

\subsection{Defining Physicality} \label{sec:physicality}

The physicality of warp metrics is established through a set of conditions, known as energy conditions, which are imposed on the stress-energy tensor components \cite{2020CQGra..37s3001K}. Among these conditions, the most frequently used and implemented within Warp Factory are:

\begin{enumerate}
    \item \textit{The Point-wise Null Energy Condition (NEC)}
    is the null (light-ray) frame's energy density observation, at all points in spacetime. For this condition to be physical, the density should be non-negative for any null observer at every point in spacetime.

    \item \textit{The Point-wise Weak Energy Condition (WEC)} is a timelike frame's energy density observation, at all points in spacetime. For this condition to be physical, the density should be non-negative for any timelike observer at every point in spacetime.

    \item \textit{The Point-wise Strong Energy Condition (SEC)} is a measure of the experienced tidal effect from the matter present acting on timelike observers. For this condition to be physical, tidal effects must be non-negative, meaning that matter should gravitate together at every point in spacetime.

    \item \textit{The Point-wise Dominant Energy Condition (DEC)} is an observation of the matter flow rate at all points in spacetime. For this condition to be physical, the velocity must be observed to be less than the speed of light.
    
\end{enumerate}

All known forms of classical matter obey these conditions. However, quantum systems can violate them for local points \cite{1995PhRvD..51.4277F}. Therefore, a softer version of the NEC, WEC, and SEC is the average of each of those conditions where each energy condition is integrated over a geodesic path instead of evaluated pointwise. In addition, there is an argument to be made to drop the SEC since dark energy and inflation appear to violate it \cite{hawking_ellis_1973}. For this analysis, however, we will evaluate the more stringent requirements of satisfaction for all energy conditions for all observers, everywhere.

Satisfaction of the energy conditions alone cannot guarantee the prospect of manmade warp drives. Any warp engineer will have to include practical requirements for the drive, such as:
\begin{itemize}
    
    \item \textit{Conservation of stress-energy $\nabla_\nu T^{\mu\nu} = 0$.} This is the conservation law of general relativity. If the stress-energy tensor is given by \eqref{einEq}, then, by construction, any metric that is continuous up to the second order will satisfy this condition per the contracted Bianchi identities \cite{2004sgig.book.....C}.

    \item \textit{No horizons of inaccessibility between the passenger volume and the external spacetime.} Warp structures should always be limited to a regime where observers can be connected to each other across the warp bubble and to the external space.

    \item \textit{No violation of causality.} Closed time-like curves are undesirable.

    \item \textit{Reasonable total energy and energy densities}  Any practical warp bubble should use an attainable amount of matter in configurations of densities that are practically achievable.
    
\end{itemize}

\subsection{Observing Stress-Energy} \label{sec:stressenergy}
The requirements of a warp drive are a description of the spacetime curvature given by the metric tensor $g_{\mu\nu}$. However, to find the energy, momenta, pressures, and shears required to generate that curvature in spacetime, the Einstein field equations must be used to compute the stress-energy tensor from the metric tensor. Einstein notion will be used for summations where the 0 component of an indexed object will be reserved for a time-like direction whereas the 1-3 components are for space-like directions. Latin indices denote summations from one to three and Greek indices denote summations from zero to three.

The components of the stress-energy tensor represent the coordinate observer's\footnote{The coordinate observers represent observers whose local metric is Minkowskian in the specified metric. For example, for the Schwarzschild metric in Schwarzschild coordinates, the coordinate observers are stationary observers at infinity.} perspective on the stress-energy throughout the space. To better understand the actual distribution of energy, momenta, pressures, and shears, local observations of these quantities must be made using observers at those same points in spacetime.

\subsubsection{Constructing Observers}
Observers in general relativity are fully specified by a timelike vector and three spatial vectors, with the timelike vector usually specified as a four-velocity. This can also be put into a tetrad formalism\footnote{Often observers are only associated with a four-velocity, however, this leaves the spatial vectors undefined. Fully defined observers with a spatial vector will have a unique tetrad. For many problems, only the four-velocity is required.} where each observer has a tetrad that is associated with their frame. The tetrad can be given as a matrix whose columns are constructed from the time vector or four-velocity $e^{\mu}_{\hat{0}}$ of the observer and its spatial vectors $e^{\mu}_{\hat{i}}$:
\begin{equation}
    e^{\mu}_{\hat{\nu}} = 
    \begin{pmatrix}
         \vrule & \vrule & \vrule & \vrule  \\ 
        e^{\mu}_{\hat{0}} & e^{\mu}_{\hat{1}} & e^{\mu}_{\hat{2}}   & e^{\mu}_{\hat{3}} \\  \vrule & \vrule & \vrule & \vrule  \\  
    \end{pmatrix}
\end{equation}
When evaluating the field equation from the metric tensor what is returned is the stress-energy as represented in the coordinate frame. Generally, we can express any frame's stress-energy tensor $T_{\hat{\mu}\hat{\nu}}$ in terms of the coordinate tensor and the frame's tetrads:
\begin{equation}
        T_{\hat{\mu}\hat{\nu}} = T_{\mu\nu} e^{\mu}_{\hat{\mu}} e^{\nu}_{\hat{\nu}}
\end{equation}
where the hats represent the observer frame. If the observer is aligned with the frame defined by the tetrad, then the observables of energy $(\rho)$, momenta $(p)$, pressures $(P)$, and shears $(\sigma)$ are directly given as:
 \begin{equation}
     \begin{split}
         \rho = T_{\hat{\mu}\hat{\nu}} n^{\hat{\mu}} n^{\hat{\nu}} = T_{\hat{0}\hat{0}} =  T^{\hat{0}\hat{0}}  \\
         p_{i} = -T_{\hat{\mu}\hat{\nu}} n^{\hat{\mu}} \gamma^{\hat{\nu}}_{i} = -T_{\hat{0}\hat{i}} =  T^{\hat{0}\hat{i}}  \\
         P_{i} = T_{\hat{\mu}\hat{\nu}} \gamma^{\hat{\mu}}_{i} \gamma^{\hat{\nu}}_{i} = T_{\hat{i}\hat{i}} = T^{\hat{i}\hat{i}} \\
         \sigma_{ij} = T_{\hat{\mu}\hat{\nu}} \gamma^{\hat{\mu}}_{i} \gamma^{\hat{\nu}}_{j} = T_{\hat{i}\hat{j}} = T^{\hat{i}\hat{j}} \\
     \end{split}
 \end{equation}
where, $i \neq j$ and, in this specific instance, $n^\mu$ and $\gamma^\mu_i$ are defined as aligned with the coordinate system, resulting in:
\begin{equation}
    n^{\hat{\mu}} = (1, 0, 0, 0), \ \ \gamma^{\hat{\mu}}_i = \begin{pmatrix}
        0 & 0 & 0 & 0 \\
        0 & 1 & 0 & 0 \\
        0 & 0 & 1 & 0 \\
        0 & 0 & 0 & 1 \\
    \end{pmatrix}
\end{equation}

\subsubsection{Eulerian Observers}
There is a definition of a particular tetrad that can be expressed in terms of the metric tensor:
\begin{equation}
    g^{\mu\nu} = e^\mu_{\hat{\mu}} e^\nu_{\hat{\nu}} \eta^{\hat{\mu} \hat{\nu}}
\end{equation}
With this tetrad, we can construct the inverse of this statement where the tetrad acts on the covariant form of the metric to return the locally-Minkowskian metric: 
\begin{equation}
     g_{\mu\nu} e^\mu_{\hat{\mu}} e^\nu_{\hat{\nu}} = \eta_{\hat{\mu} \hat{\nu}}
\end{equation}
Finding the observer tetrad is then simply solving this equation given the metric tensor. This solution is not unique unless the temporal and spatial orientation is first defined, which depends on the selected observer.

A special set of observers is the set of \textit{Eulerian Observers} which is defined as having a tetrad with a four-velocity that is orthogonal to the spatial hypersurface (not moving in space over time). Using 3+1 formalism the foliation of the spacetime is constructed as \cite{2007gr.qc.....3035G}:
\begin{equation}
    g_{\mu\nu} = 
    \begin{pmatrix}
    -\alpha^2 + \beta_i \beta^i & \beta_i \\
    \beta_i & \gamma_{ij}
    \end{pmatrix}
\end{equation}
where $\alpha$ represents the lapse rate and where the spatial metric $\gamma_{ij}$ raises the shift vector index $\beta_i$ as:
\begin{equation}
    \beta^i = \gamma^{ij}\beta_j
\end{equation}
We can then decompose any metric defined in the standard form into its 3+1 components:
\begin{equation}
    \begin{split}
    &\beta_i = g_{0i} \\
    &\gamma_{ij} = g_{ij} \\
    &\alpha = \sqrt{g^{0i}g_{0i} - g_{00}} \\
    \end{split}
\end{equation}
This formalism allows for a convenient definition of the four-velocity which is orthogonal to the spatial hypersurface:
\begin{equation}
    e^\mu_{\hat{0}} 
 = n^\mu = \frac{1}{\alpha}\left(1,-\beta^1,-\beta^2,-\beta^3 \right), \ \ n_\mu = \left(-\alpha,0,0,0\right)
 \label{eq:eulerianVelocity}
\end{equation}
and the spatial terms of $e^{\mu}_{\hat{i}}$ can be selected freely assuming they are orthogonal to the time vector and each other. In Warp Factory we define the spatial vectors by selecting that $\hat{x}$ is orthogonal to the coordinate $y$ and $z$ surface and the remaining spatial vectors being orthogonal to each other (full solution and vector construction are detailed in \ref{apx:euleriantransformation})\footnote{This choice of alignment is arbitrary but needed to provide a unique solution.}. With this setup, we have fully defined an Eulerian tetrad with a selected spatial orientation which can be solved symbolically from only the covariant metric.

\section{Methods of Warp Factory} \label{sec:methods}
In Warp Factory, each of the 10 independent metric components ($g_{\mu\nu}$) can be specified at each point on the 4D spacetime grid. For correct evaluations, the metric is required to have a Minkowskian signature of (-,+,+,+) at all points.

\subsection{Evaluating the Field Equation}\label{sec:metricEFEMethod}
The stress-energy tensor is determined from the metric through the field equations:
\begin{equation}
   T_{\mu\nu} = \frac{c^4}{8 \pi G} \left( R_{\mu\nu}-\frac{1}{2}R g_{\mu\nu} \right) 
   \label{einEq}
\end{equation}
In Warp Factory, the Ricci tensor $R_{\mu\nu}$ is calculated directly in terms of the metric derivatives:
\begin{equation}
    \begin{split}
        R_{\mu \nu} & =-\frac{1}{2} \sum_{\alpha, \beta=0}^3\left(\frac{\partial^2 g_{\mu \nu}}{\partial x^\alpha \partial x^\beta}+\frac{\partial^2 g_{\alpha \beta}}{\partial x^\mu \partial x^\nu}-\frac{\partial^2 g_{\mu \beta}}{\partial x^\nu \partial x^\alpha}-\frac{\partial^2 g_{\nu \beta}}{\partial x^\mu \partial x^\alpha}\right) g^{\alpha \beta} \\
        & +\frac{1}{2} \sum_{\alpha, \beta, \gamma, \delta=0}^3\left(\frac{1}{2} \frac{\partial g_{\alpha \gamma}}{\partial x^\mu} \frac{\partial g_{\beta \delta}}{\partial x^\nu}+\frac{\partial g_{\mu \gamma}}{\partial x^\alpha} \frac{\partial g_{\nu \delta}}{\partial x^\beta}-\frac{\partial g_{\mu \gamma}}{\partial x^\alpha} \frac{\partial g_{\nu \beta}}{\partial x^\delta}\right) g^{\alpha \beta} g^{\gamma \delta} \\
        & -\frac{1}{4} \sum_{\alpha, \beta, \gamma, \delta=0}^3\left(\frac{\partial g_{\nu \gamma}}{\partial x^\mu}+\frac{\partial g_{\mu \gamma}}{\partial x^\nu}-\frac{\partial g_{\mu \nu}}{\partial x^\gamma}\right)\left(2 \frac{\partial g_{\beta \delta}}{\partial x^\alpha}-\frac{\partial g_{\alpha \beta}}{\partial x^\delta}\right) g^{\alpha \beta} g^{\gamma \delta}
    \end{split}\label{eq:Ricci}
\end{equation}
 The Ricci scalar $R$ is:
\begin{equation}
    R = R_{\mu\nu}g^{\mu\nu}
    \label{eq:RicciScalar}
\end{equation}
The metric derivatives are found using a fourth-order central finite difference method \cite{shibata_2016} with a 1-meter grid spacing. The boundary points of the metric are cropped in the evaluations. See appendix \ref{apx:errors} for details on accuracy.

For the computation of $T^{\hat{\mu}\hat{\nu}}$, tetrads are selected to fulfill the requirements of \eqref{eq:eulerianVelocity}. A full solution in terms of the covariant metric terms are detailed in \ref{apx:euleriantransformation}. This alternative approach is used within Warp Factory as the metrics are not strictly constructed in 3+1 formalism\footnote{The metrics created and evaluated with this method must be able to be constructed in 3+1. The tetrad creation is equivalent}.

\subsection{Evaluating Physicality} \label{sec:evalphysicality}
In Warp Factory, to establish if a solution is physical, we evaluate the standard energy conditions (NEC, WEC, SEC, and DEC) for all points in discretized spacetime. All evaluations of the energy conditions require a contraction of the stress-energy tensor with an observer vector field that is either null or timelike. Numerically, this process involves sampling observer velocities and contracting them with the stress-energy tensor at all spacetime points on the coordinate grid. The stress-energy tensor and the energy conditions will be evaluated in the tetrad frame, $T^{\hat{\mu}\hat{\nu}}$, which, at all points, is associated with a local Minkowskian metric. Each observer is defined by a four-velocity that can be either null or time-like. The evaluation of energy conditions does not require a specification of the spatial vectors of the tetrad if the tensor is transformed into any Eulerian frame. It can simply be represented as the contraction of the stress-energy tensor with four-velocities, so we drop the strict tetrad description here for convenience. 

\subsubsection{Observer Vectors and Vector Fields}
The null vector for a given observer is $k^{\hat{\mu}}$, which, in a Cartesian locally-Minkowskian space, is defined as:
\begin{equation}
    k^{\hat{\mu}} = (1,\vec{u})
\end{equation}
where $\vec{u}$ is some vector representing the spatial velocity of an observer that satisfies the condition of $|\vec{u}| = 1$. For the set of null observers $\textbf{O}_\textbf{n}$, the direction of the spatial velocity is sampled from a set of vectors that map to evenly distributed points on a sphere, which, in a spherical coordinate representation, result in the set of four-velocities that map to each observer as:
\begin{equation}
    \textbf{O}_\textbf{n} = \begin{bmatrix}
     O^1 \rightarrow k^{\hat{\mu}} = (1, \ \sin(\theta_1)\cos(\phi_1), \ \sin(\theta_1)\sin(\phi_1), \ \cos(\theta_1)) \\
     \vdots \\
     O^n \rightarrow k^{\hat{\mu}} = (1, \ \sin(\theta_n)\cos(\phi_n), \ \sin(\theta_n)\sin(\phi_n), \ \cos(\theta_n))
    \end{bmatrix}
\end{equation}

The resulting vectors for each observer $O^n$ are then normalized. For timelike vectors, $V^{\hat{\mu}}$, the four-velocity is defined in the same manner but $|\vec{u}|$ is additionally scaled from 0 to 1 for a specified number of timelike samples for each of the spatial velocity direction samples. The timelike observers, $\textbf{O}_\textbf{t}$, are constructed as:

\begin{equation}
    \textbf{O}_\textbf{t} = \begin{bmatrix}
     O^1 \rightarrow V^{\hat{\mu}} = (1, \ \sin(\theta_1)\cos(\phi_1)s_1, \ \sin(\theta_1)\sin(\phi_1)s_1, \ \cos(\theta_1)s_1) \\
     \vdots \\
     O^n \rightarrow V^{\hat{\mu}} = (1, \ \sin(\theta_n)\cos(\phi_n)s_n, \ \sin(\theta_n)\sin(\phi_n)s_n, \ \cos(\theta_n)s_n)
    \end{bmatrix}
\end{equation}
where $0<s_n<1$. All vectors are then normalized. 

Since the metric, $g_{\hat{\mu}\hat{\nu}}$, is locally-Minkowskian everywhere, the same vector can be contracted at all spacetime points $X$ for $T_{\hat{\mu}\hat{\nu}}(X)$. This means that $k^{\hat{\mu}}$ and $V^{\hat{\mu}}$ can be treated as uniform vector fields since their values do not vary over the spacetime.

\subsubsection{Energy Conditions}
The Null Energy Condition (NEC) is given by the contraction of stress-energy tensor with all observer null vector fields:
\begin{equation}
    \Xi_{N}(X) = T_{\hat{\mu}\hat{\nu}}(X) k^{\hat{\mu}} k^{\hat{\nu}} \ge 0 \ \ \forall \ \  k^{\hat{\mu}}
\end{equation}
Similarly, the Weak Energy Condition (WEC) is a contraction of the stress-energy tensor with all observer timelike vector fields:
\begin{equation}
    \Xi_{W}(X) = T_{\hat{\mu}\hat{\nu}}(X) V^{\hat{\mu}} V^{\hat{\nu}} \ge 0 \ \ \forall \ \  V^{\hat{\mu}}
\end{equation}
The Strong Energy Condition (SEC) is found by the equation:
\begin{equation}
    \Xi_{S}(X) = \left(T_{\hat{\mu}\hat{\nu}}(X) - \frac{1}{2} T(X) \eta_{\hat{\mu}\hat{\nu}} \right)V^{\hat{\mu}} V^{\hat{\nu}} \ge 0 \ \ \forall \ \  V^{\hat{\mu}}
\end{equation}
Lastly, the Dominant Energy Condition (DEC) is satisfied when both the WEC is greater than zero and when the contracted mixed tensor with a timelike vector field is less than zero. The latter part of the DEC is given by:
\begin{equation}
    \Upsilon^{\hat{\mu}}(X) = -T^{\hat{\mu}}_{\ \ \hat{\nu}}(X) V^{\hat{\nu}}
\end{equation}
where this new vector field $\Upsilon^{\hat{\mu}}\left(X\right)$ must be future pointing and either timelike or null. This is found by the equation:
\begin{equation}
    \xi_D(X) = \eta_{\hat{\mu}\hat{\nu}} \Upsilon^{\hat{\mu}}(X) \Upsilon^{\hat{\nu}}(X) \le 0 \ \ \forall \ \  V^{\hat{\mu}}
\end{equation}
For easy comparison to other energy conditions, the square root of this value is taken while flipping the overall sign:
\begin{equation}
    \Xi_{D}(X) =  -\textrm{sign}(\xi_D(X))\sqrt{|\xi_D(X)|}
\end{equation}
Therefore, just like the other energy conditions, when the DEC is evaluated, a negative result represents the violation of the condition. In Warp Factory, only the evaluation of $\Xi_{D}(X)$ is returned as the dominant energy condition.

For each energy condition, the minimum value across all observers is taken for each point in the spacetime grid. These minimum values are what is plotted in the energy condition evaluations presented in Section \ref{sec:metricEval}.

\subsection{Momentum Flowlines} \label{sec:momentum}
 Warp metrics generate terms beyond just the energy density. One unique feature that often exists in warp drive spacetimes is momentum flux $p_i$, which describes the motion of the energy density.

 This flow can be visualized by constructing a set of flowlines that are built from the vector field of the momentum density, given by $\boldsymbol{\Omega}$:
\begin{equation}
    \Omega_i(X) = \left(p_1(X), p_2(X), p_3(X)\right)
\end{equation}

Each of the flowlines is found by propagating a virtual particle from a starting point through the vector field for a given number of steps. It should be noted that while this provides insight into the structure of the momentum density, these lines are not those of matter moving along geodesic paths.

\subsection{Metric Scalar Analysis}\label{sec:metricScalarMethod}

The metric tensor can also be understood by inspecting its associated geometric scalars, which are constructed from the metric tensor by projections onto a timelike vector field $U^\mu$. Any timelike vector field can be used, but the specific vector field assumed in Warp Factory is that of a four-velocity orthogonal to the spatial hypersurface as given by $U^\mu = n^\mu$.
\subsubsection{Expansion Scalar}
The expansion scalar describes how the volume of a region changes. We define this scalar $\theta$ as:
\begin{equation}
   \theta = g^{\mu\nu}\theta_{\mu\nu}
\end{equation}
where $\theta_{\mu\nu}$ is the stress tensor, which is computed using the projection tensor $P^\alpha_\mu$ that operates on the covariant derivative of the vector field $U^\mu$ as\footnote{Note: $X_{(ab)} = X_{ab} + X_{ba}$ and $X_{[ab]} = X_{ab} - X_{ba}$.}
\begin{equation}
    \theta_{\mu\nu} = P^\alpha_\mu P^\beta_\nu \nabla_{(\alpha} U_{\beta)}
\end{equation}
and the projection tensor is given by:
\begin{equation}
    P_{\mu\nu} = g_{\mu\nu} + U_\mu U_\nu
\end{equation}

\subsubsection{Shear Scalar}
The shear scalar describes the relative stretching while preserving the object's volume. The shear scalar $\sigma^2$ is defined as:
\begin{equation}
    \sigma^2 = \frac{1}{2} \sigma_{\mu\nu} \sigma^{\mu\nu}
\end{equation}
where $\sigma_{\mu\nu}$ is the shear tensor, which is similarly computed using the stress tensor along with the projection tensor $P_{\mu\nu}$, shear tensor $\theta_{\mu\nu}$, and shear scalar $\theta$ as:
\begin{equation}
   \sigma_{\mu\nu} = \theta_{\mu\nu} - \frac{\theta}{3} P_{\mu\nu}
\end{equation}

\subsubsection{Vorticity Scalar}
The vorticity scalar is the magnitude of the twist in spacetime, given by:
\begin{equation}
   \omega^2 = \frac{1}{2}\omega_{\mu\nu}\omega^{\mu\nu}
\end{equation}
Where $\omega_{\mu\nu}$ is the vorticity tensor, which is computed using the projection tensor as
\begin{equation}
    \omega_{\mu\nu} = P^\alpha_\mu P^\beta_\nu \nabla_{[\alpha} U_{\beta]}.
\end{equation}
Given a vector field that is orthogonal to the spatial hypersurface, the vorticity scalar will always be zero, but this is not generally true for an arbitrary vector field.

\clearpage
\section{Metric Evaluations}\label{sec:metricEval}
The numerical approach that Warp Factory enables provides greater insight into the details of warp spacetimes. As a demonstration, we will consider several popular metrics discussed in the literature.

\subsection{Alcubierre Metric}
Alcubierre introduced the first general relativistic warp drive in 1994 \cite{1994CQGra..11L..73A}. By adding a localized and spherical shift to a Minkowski spacetime, one obtains a flat passenger volume moving along a pre-defined curve. The Alcubierre metric is given as:
\begin{equation}
    \mathrm{d}s^2 = -d t^2+\left(d x-v_s f\left(r_s\right) d t\right)^2+d y^2+d z^2
\end{equation}
where
\begin{equation}
\begin{split}
    &f\left(r_s\right)=\frac{\tanh \left(\sigma\left(r_s+R\right)\right)-\tanh \left(\sigma\left(r_s-R\right)\right)}{2 \tanh (\sigma R)} \\
    &r_s = \sqrt{(x-x_0-v_s t)^2+\left(y-y_0\right)^2+\left(z-z_0\right)^2}
    \end{split}
\end{equation}
To analyze the metric, the commonly used Galilean transformation is applied:
\begin{equation}
    \begin{split}
        &x^\prime = x-v_s t \\
        &t^\prime = t
    \end{split}   
\end{equation}
This makes the metric time-independent and amounts to an addition of the shift vector:
\begin{equation}
    \beta^{\prime}_x = \beta_x+v_s
\end{equation}

\begin{figure}[hbt!]
\centering
\includegraphics[width=\textwidth]{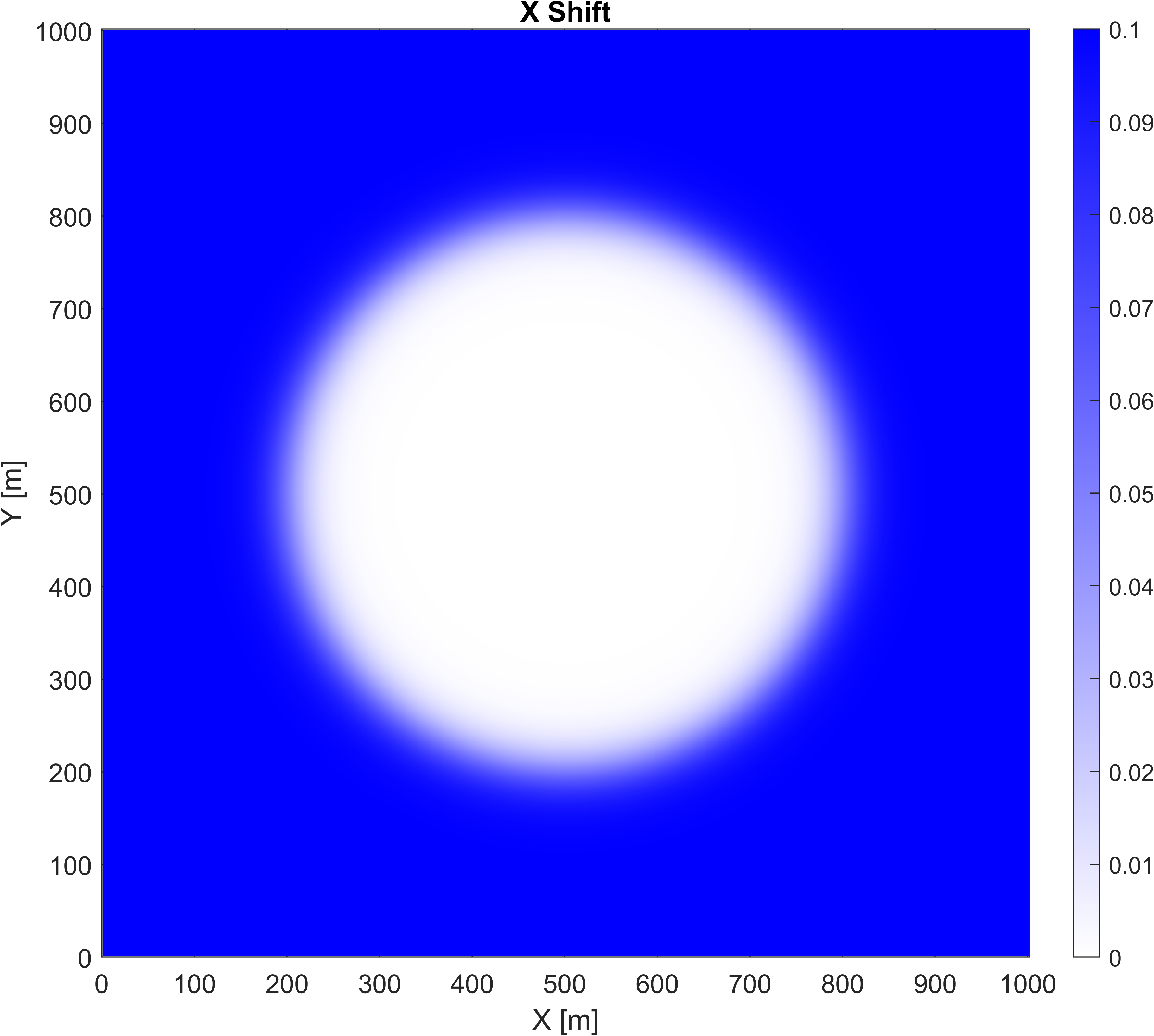}
\caption{Alcubierre metric comoving shift vector for the X-direction. The motion of the bubble is in the +X direction. The slice is for $z=z_0$, $v_s=0.1$ c, $R=300$ m, $\sigma=0.015$ $\text{m}^{-1}$, $x_0=y_0=z_0=503$ m.}\label{fig:AM_components}
\end{figure}

\begin{figure}[hbt!]
\centering
\includegraphics[width=\textwidth]{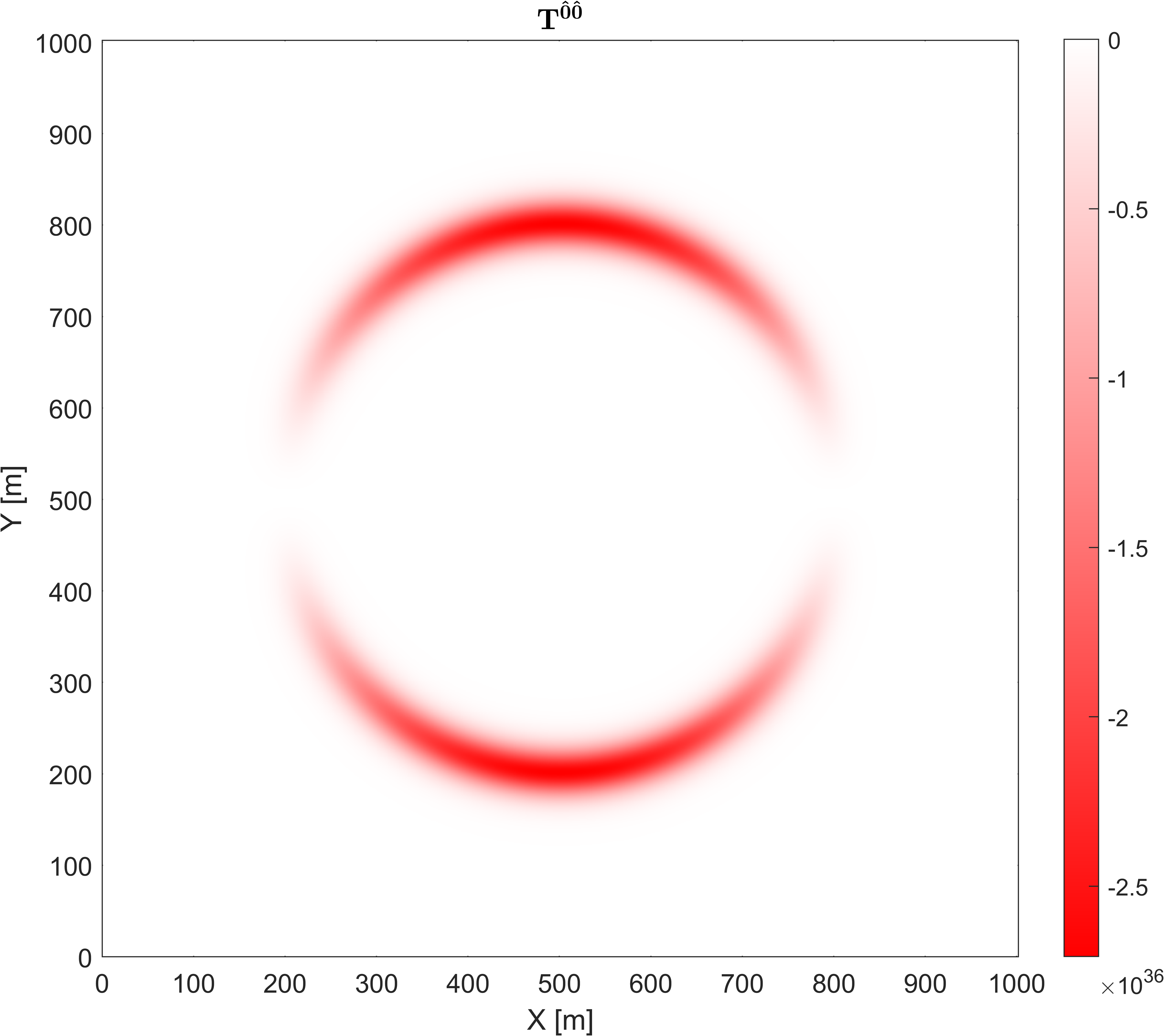}
\caption{Energy density for the Alcubierre metric. The motion of the bubble is in the +X direction. The slice is for $z=z_0$, $v_s=0.1$ c, $R=300$ m, $\sigma=0.015$ $\text{m}^{-1}$, $x_0=y_0=z_0=503$ m. Units are in J/m$^3$.}\label{fig:AM_Eden}
\end{figure}

\begin{figure}[hbt!]
\centering
\includegraphics[width=\textwidth]{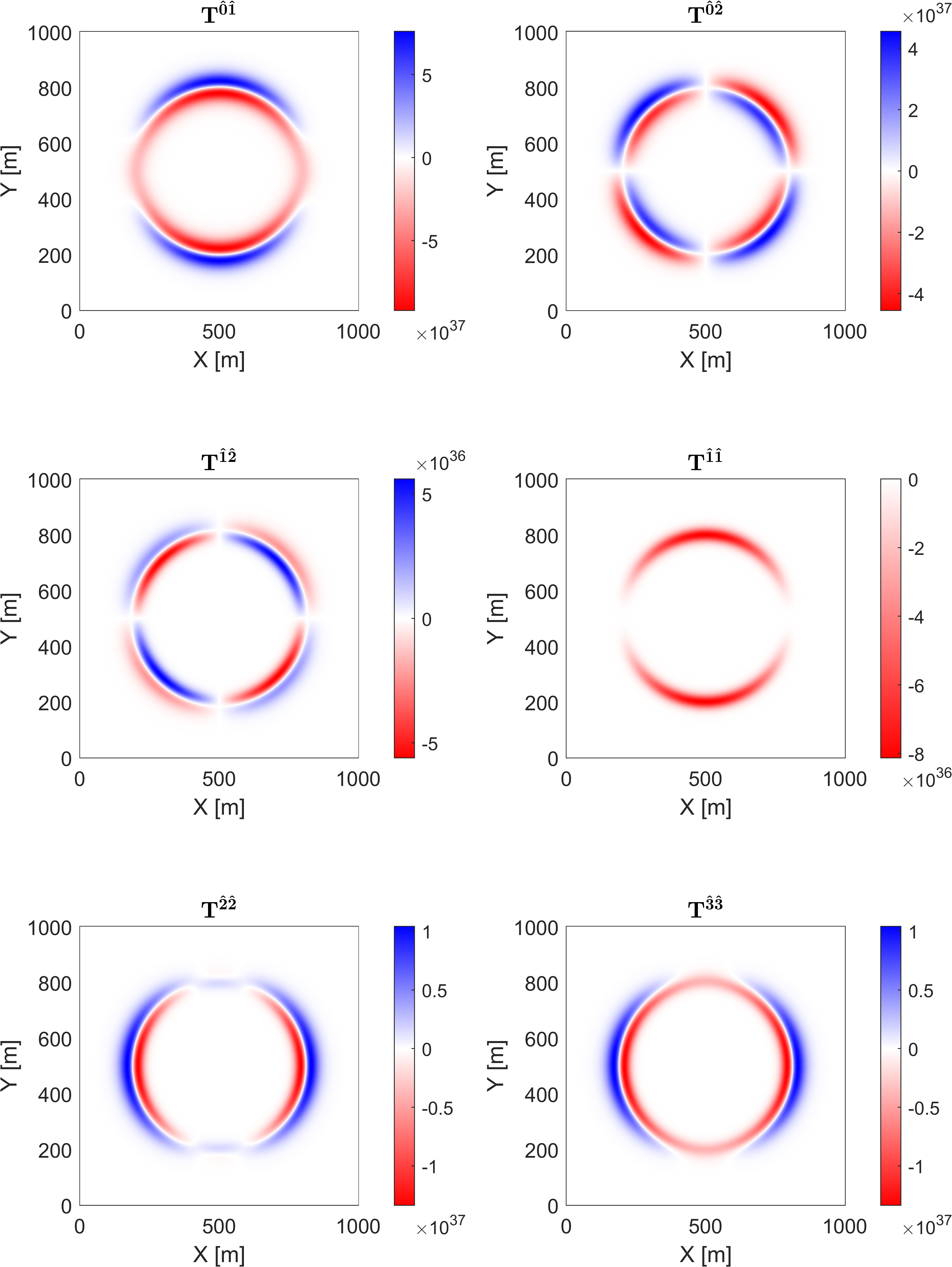}
\caption{The stress-energy tensor for the Alcubierre metric. The $T^{\hat{0}\hat{0}}$ component is shown in Figure \ref{fig:AM_Eden}.  The motion of the bubble is in the +X direction. The slice is for $z=z_0$, $v_s=0.1$ c, $R=300$ m, $\sigma=0.015$ $\text{m}^{-1}$, $x_0=y_0=z_0=503$ m. Units are in J/m$^3$.}\label{fig:AM_Etensor}
\end{figure}

\begin{figure}[hbt!]
\centering
\includegraphics[width=0.95\textwidth]{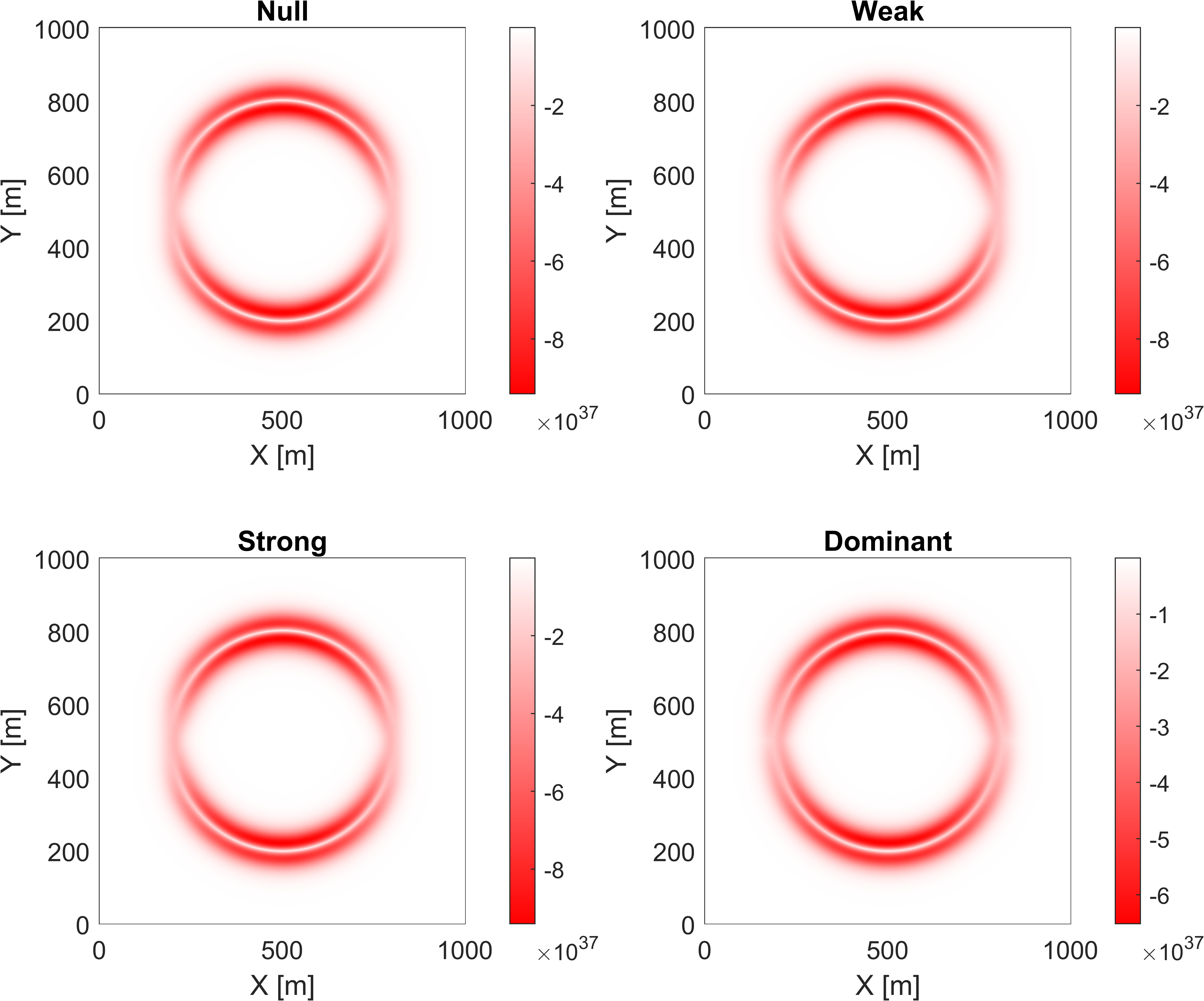}
\caption{Energy condition evaluation for the Alcubierre metric. The motion of the bubble is in the +X. The slice is for $z=z_0$, $v_s=0.1$ c, $R=300$ m, $\sigma=0.015$ $\text{m}^{-1}$, $x_0=y_0=z_0=503$ m, $n_{\rm{observers}}=1000$. The minimum value among all observers is shown. Units are in J/m$^3$.}\label{fig:AM_Econd}
\end{figure}

\begin{figure}[hbt!]
\centering
\includegraphics[width=0.95\textwidth]{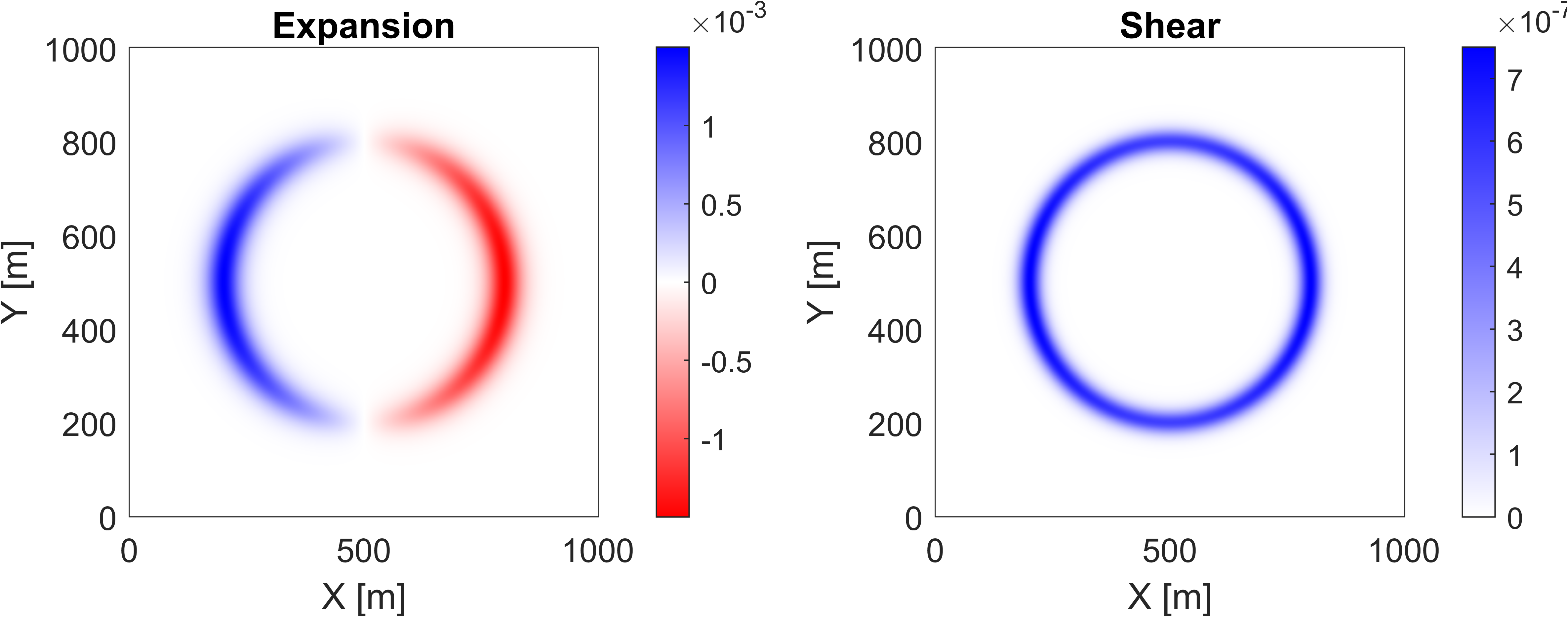}
\caption{Expansion and shear scalars for the Alcubierre metric. The motion of the bubble is in the +X. The slice is for $z=z_0$, $v_s=0.1$ c, $R=300$ m, $\sigma=0.015$ $\text{m}^{-1}$, $x_0=y_0=z_0=503$ m. }\label{fig:AM_scalars}
\end{figure}

The metric being in a comoving frame removes the necessity for time derivatives and reduces the computation required. The Alcubierre metric violates all energy conditions within the bubble region, as it necessitates negative energy density, as Alcubierre demonstrated. This fact is confirmed through the analysis conducted using Warp Factory. The metric solved used the following parameters: $v_s=0.1$ c, $R=300$ m, $\sigma=0.015$ $\text{m}^{-1}$, $x_0=y_0=z_0=503$ m. The ratio of R to sigma was chosen so that there is a substantial region of flat space inside of the passenger area and so that the derivatives of the shift were not extremely large across the bubble boundary. In addition, R was chosen sufficiently large to have high accuracy when evaluating the stress-energy tensor and the energy conditions at a 1-meter grid spacing. This creates a metric which is shown in Figure \ref{fig:AM_components}. The resulting stress-energy tensor is shown in Figure \ref{fig:AM_Eden}, which depicts the energy density, and Figure \ref{fig:AM_Etensor} which presents the complete stress-energy components. Finally, Figure \ref{fig:AM_Econd} showcases the energy conditions, and Figure \ref{fig:AM_scalars} shows the metric scalars.


\clearpage

\subsection{Van Den Broeck Metric}

\begin{figure}[hbt!]
\centering
\includegraphics[width=\textwidth]{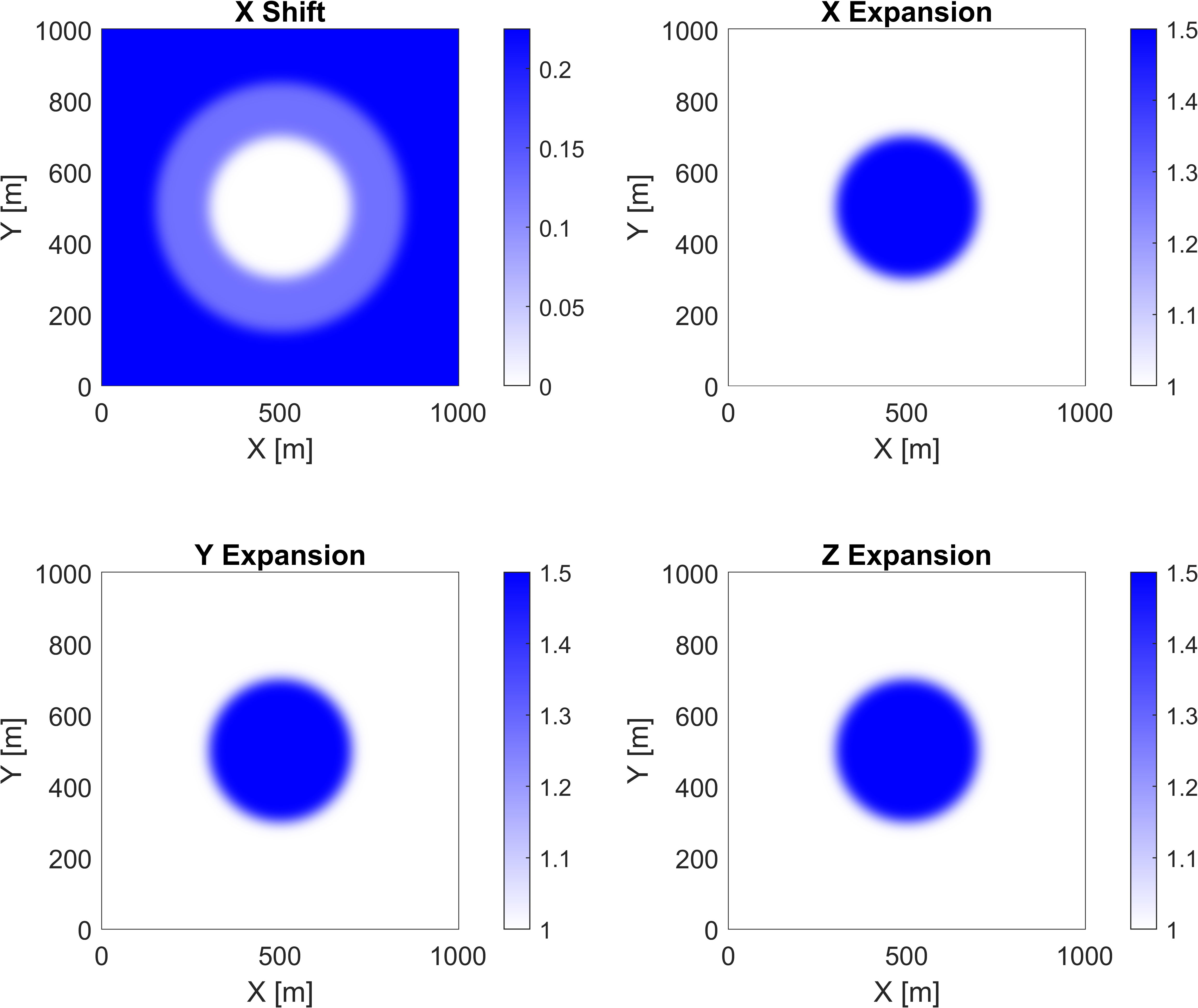}
\caption{Metric components for the comoving Van Den Broeck metric. The Van Den Broeck metric uses a shift vector in the X direction and expansion in $\gamma_{ii}$ terms. The motion of the bubble is in the +X direction. The slice is for $z=z_0$, $v_s=0.1$ c, $\alpha=0.5$, $R=350$ m, $\tilde{R}=200$ m, $\Delta=\tilde{\Delta}=40$ m, $x_0=y_0=z_0=503$ m.}\label{fig:VDM_metric}
\end{figure}

Shortly after Alcubierre's original paper, Van Den Broeck formulated his own variation of the Alcubierre metric. His approach involved employing two concentric regions of varying shift-vector and expansion terms. The objective was to diminish the need for negative energy through the increase in the volume of the passenger region. This formulation, as defined by Van Den Broeck's initial publication, is given by:
\begin{equation}
    \mathrm{d}s^2 =-d t^2+B^2\left(r_s\right)\left[\left(d x-v_s(t) f\left(r_s\right) d t\right)^2+d y^2+d z^2\right]
\end{equation}
where $f(r_s)$ is:
\begin{equation}
    \begin{split}
    f(r_s) =  1 & \text { for } \quad r_s<R \\
    0<f\left(r_s\right) \leq 1 & \text { for } \quad R \leq r_s<R+\Delta \\
    f(r_s) = 0 & \text { for } \quad R+\Delta \leq r_s
    \end{split}
\end{equation}
and $B(r_s)$ is:
\begin{equation}
   \begin{split}
     B(r_s) = 1+\alpha & \text { for } \quad r_s<\tilde{R} \\
    1<B\left(r_s\right) \leq 1+\alpha & \text { for } \quad \tilde{R} \leq r_s<\tilde{R}+\tilde{\Delta} \\
     B(r_s) = 1 & \text { for } \quad \tilde{R}+\tilde{\Delta} \leq r_s
    \end{split}
\end{equation}
Here $\tilde{R}$ is the inner region radius, $R$ is the outer radius of the warp bubble, and $\tilde{\Delta}$ and $\Delta$ are the thickness of the transition regions. As in the Alcubierre metric analysis, the Galilean transformation to a constant velocity comoving frame will be used. The parameters used are $v_s=0.1$ c, $\alpha=0.5$, $R=350$ m, $\tilde{R}=200$ m, $\Delta=\tilde{\Delta}=40$ m, $x_0=y_0=z_0=503$ m. The metric component cross-sections are shown in Figure \ref{fig:VDM_metric}.

\begin{figure}[hbt!]
\centering
\includegraphics[width=\textwidth]{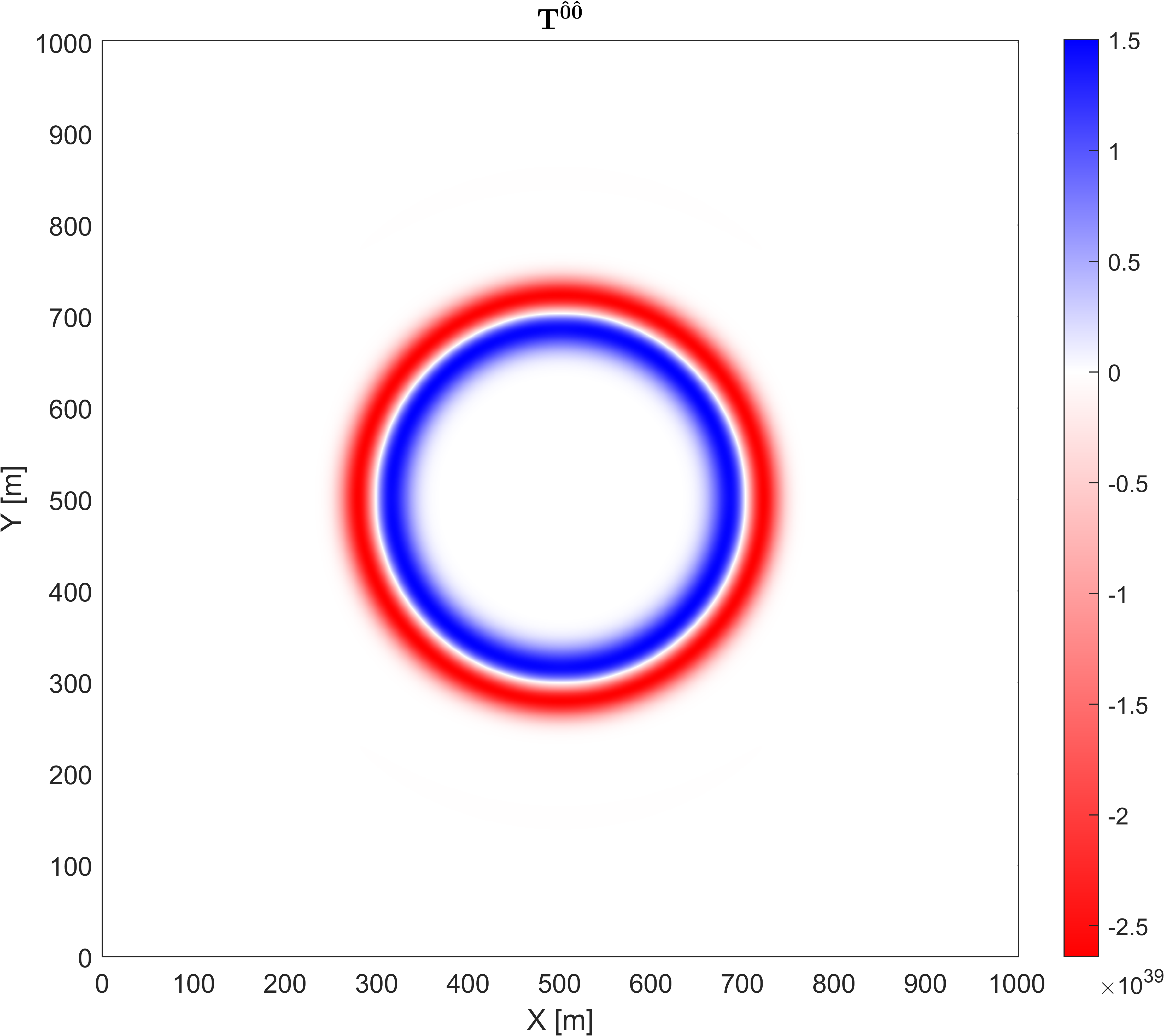}
\caption{Energy density for the Van Den Broeck metric. The motion of the bubble is in the +X direction. The slice is for $z=z_0$, $v_s=0.1$ c, $\alpha=0.5$, $R=350$ m, $\tilde{R}=200$ m, $\Delta=\tilde{\Delta}=40$ m, $x_0=y_0=z_0=503$ m. Units are in J/m$^3$.}\label{fig:VDM_Eden}
\end{figure}

\begin{figure}[hbt!]
\centering
\includegraphics[width=\textwidth]{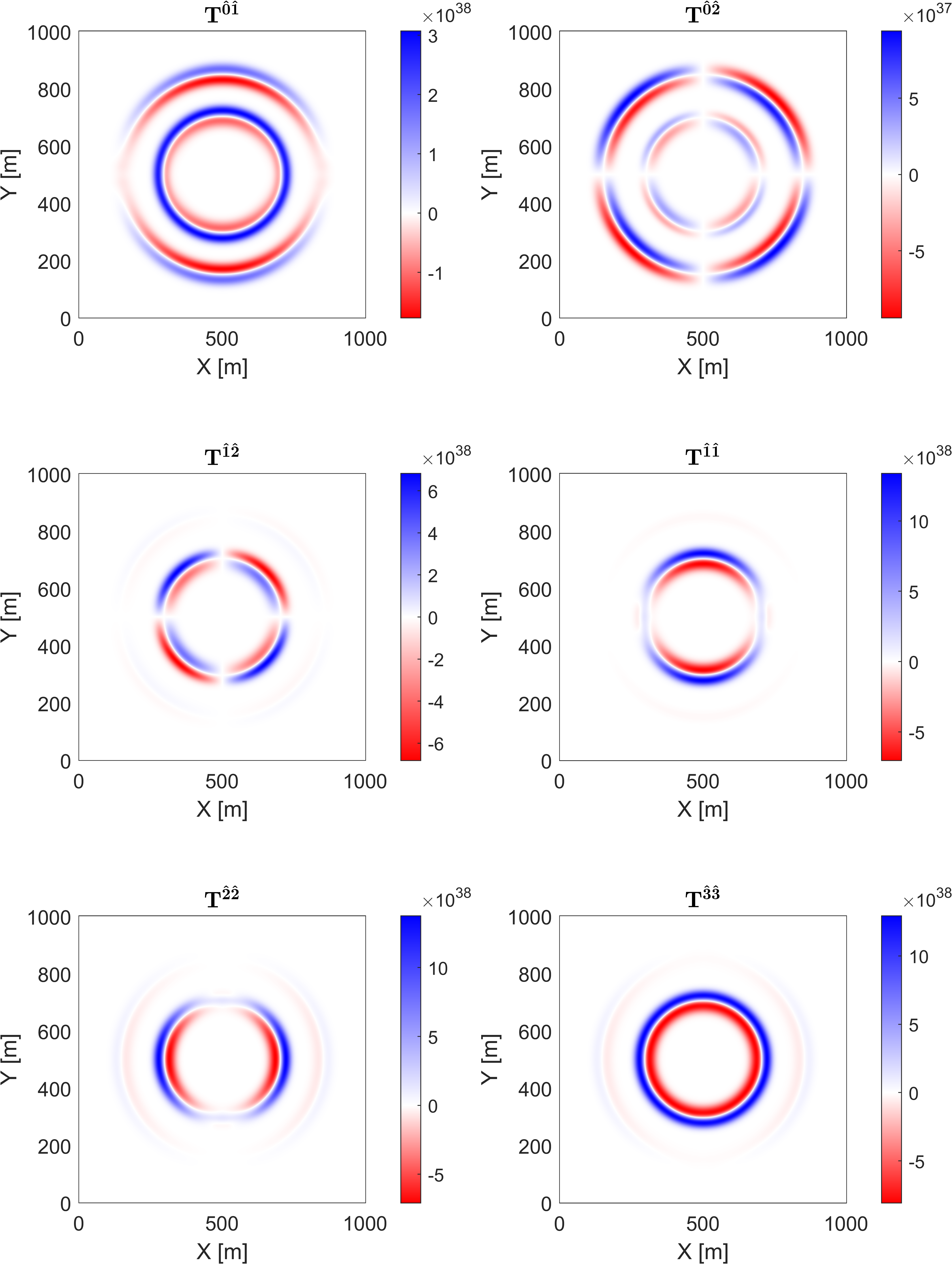}
\caption{Stress-energy tensor for the Van Den Broeck metric. The $T^{\hat{0}\hat{0}}$ component is shown in Figure \ref{fig:VDM_Eden}. The motion of the bubble is in the +X direction. The slice is for $z=z_0$, $v_s=0.1$ c, $\alpha=0.5$, $R=350$ m, $\tilde{R}=200$ m, $\Delta=\tilde{\Delta}=40$ m, $x_0=y_0=z_0=503$ m. Units are in J/m$^3$.}\label{fig:VDM_Etensor}
\end{figure}

\begin{figure}[hbt!]
\centering
\includegraphics[width=0.95\textwidth]{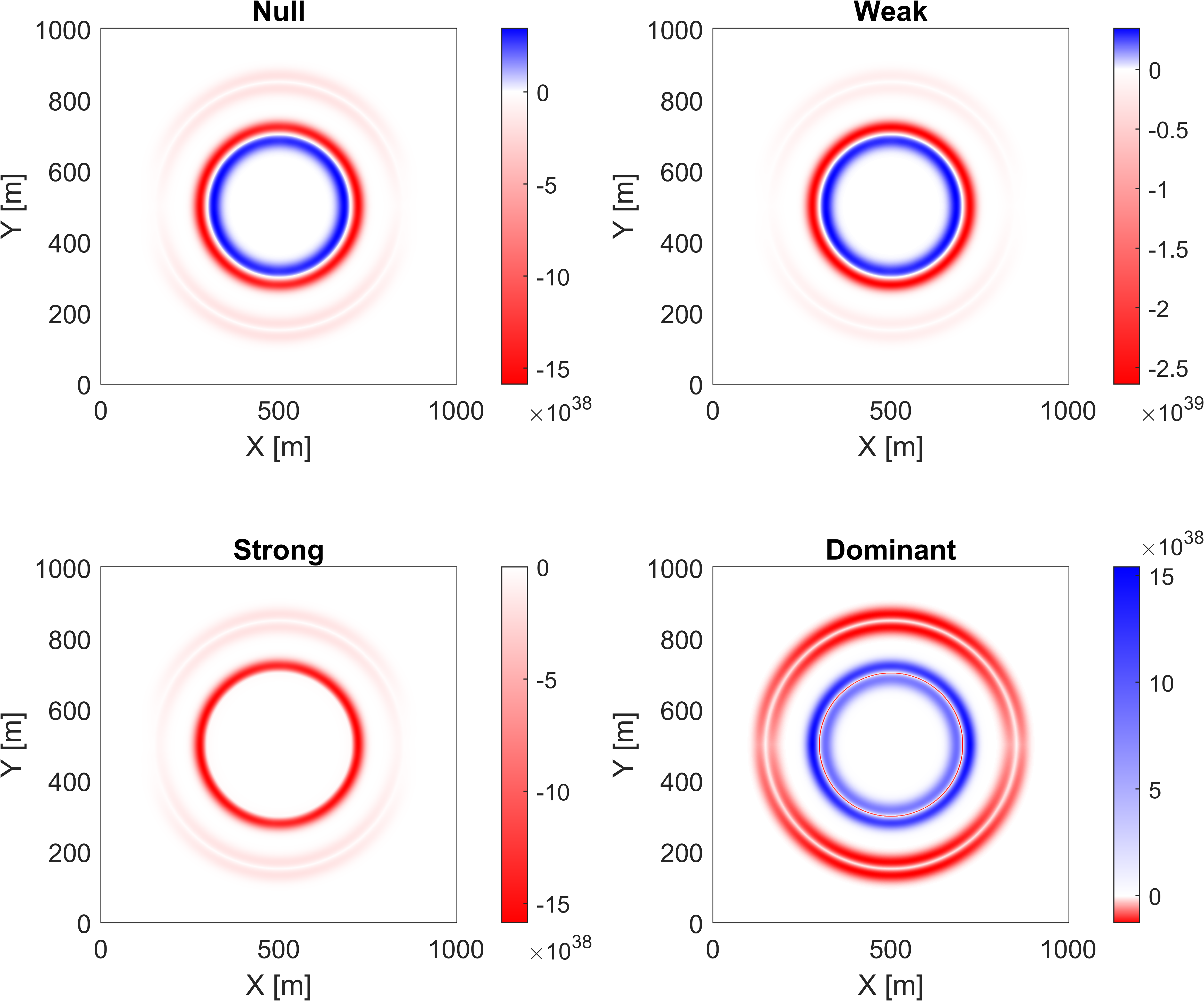}
\caption{Energy conditions for the Van Den Broeck metric. The motion of the bubble is in the +X direction. The slice is for $z=z_0$, $v_s=0.1$ c, $\alpha=0.5$, $R=350$ m, $\tilde{R}=200$ m, $\Delta=\tilde{\Delta}=40$ m, $x_0=y_0=z_0=503$ m, $n_{\rm{observers}}=1000$. The minimum value among all observers is shown which can be positive. Units are in J/m$^3$. }\label{fig:VDM_Econd}
\end{figure}

\begin{figure}[hbt!]
\centering
\includegraphics[width=0.95\textwidth]{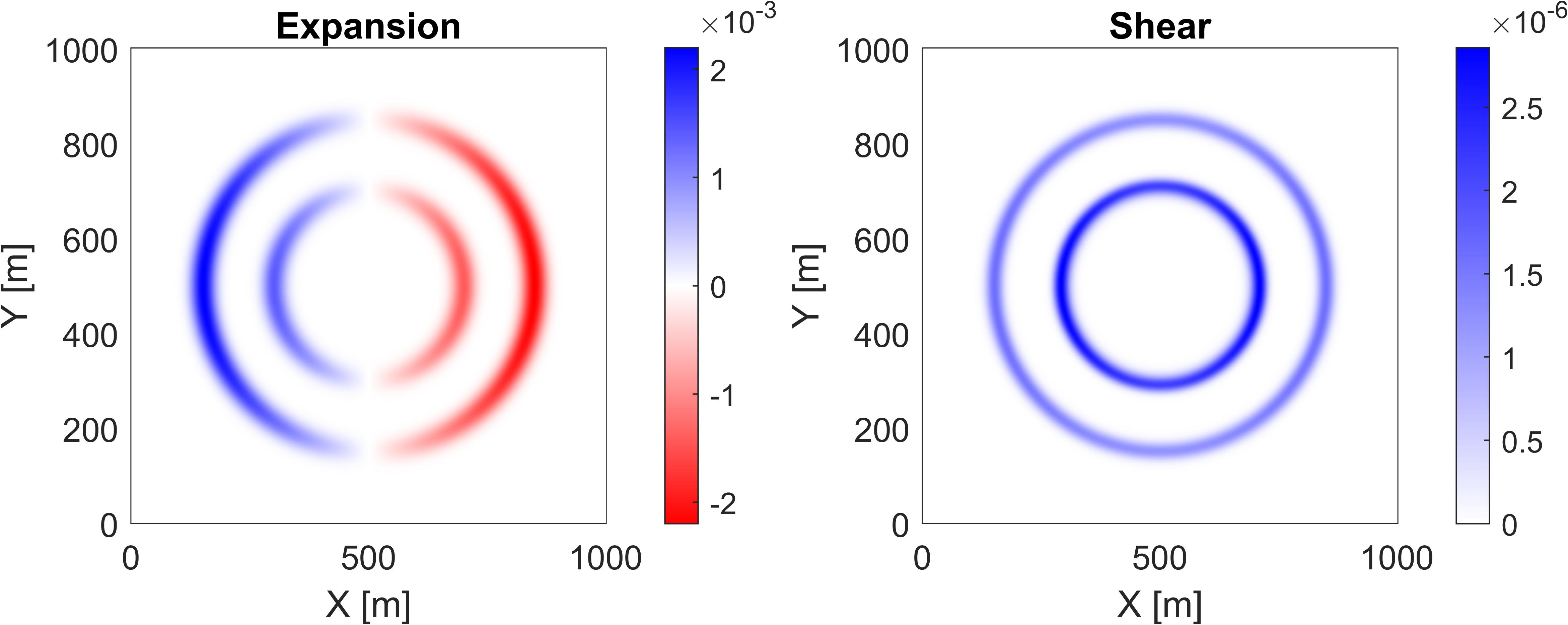}
\caption{The expansion and shear scalars for the Van Den Broeck metric. The motion of the bubble is in the +X direction. The slice is for $z=z_0$, $v_s=0.1$ c, $\alpha=0.5$, $R=350$ m, $\tilde{R}=200$ m, $\Delta=\tilde{\Delta}=40$ m, $x_0=y_0=z_0=503$ m.}\label{fig:VDM_scalars}
\end{figure}

Figure \ref{fig:VDM_Eden} illustrates the energy density profile, which shares a resemblance to Alcubierre but features a positive region within the transition into the passenger volume as it expands. Similarly, Figure \ref{fig:VDM_Econd} demonstrates that this metric also violates energy conditions, similar to Alcubierre, but with rings and a non-violating region where the positive energy exists. The presence of negative energy in the Eulerian frame indicates the existence of a violation, but understanding the distinct rings of violation requires consideration of the remaining stress-energy tensor components, depicted in Figure \ref{fig:VDM_Etensor}. The interplay between pressure and momentum density, particularly in areas of low energy density, gives rise to the primary regions of violation. Lastly, the Van Den Broeck metric scalars, shown in Figure \ref{fig:VDM_scalars}, have many similar features to Alcubierre, just divided along its concentric shell regions.

Van Den Broeck's approach does not address the key issue which creates the energy condition violations in the Alcubierre metric, but it may provide a unique method to reduce the total violation amount per passenger volume, thus creating a more efficient warp drive through its use of volume expansion.


\clearpage

\subsection{Bobrick-Martire Modified Time Metric}
\begin{figure}[hbt!]
\centering
\includegraphics[width=\textwidth]{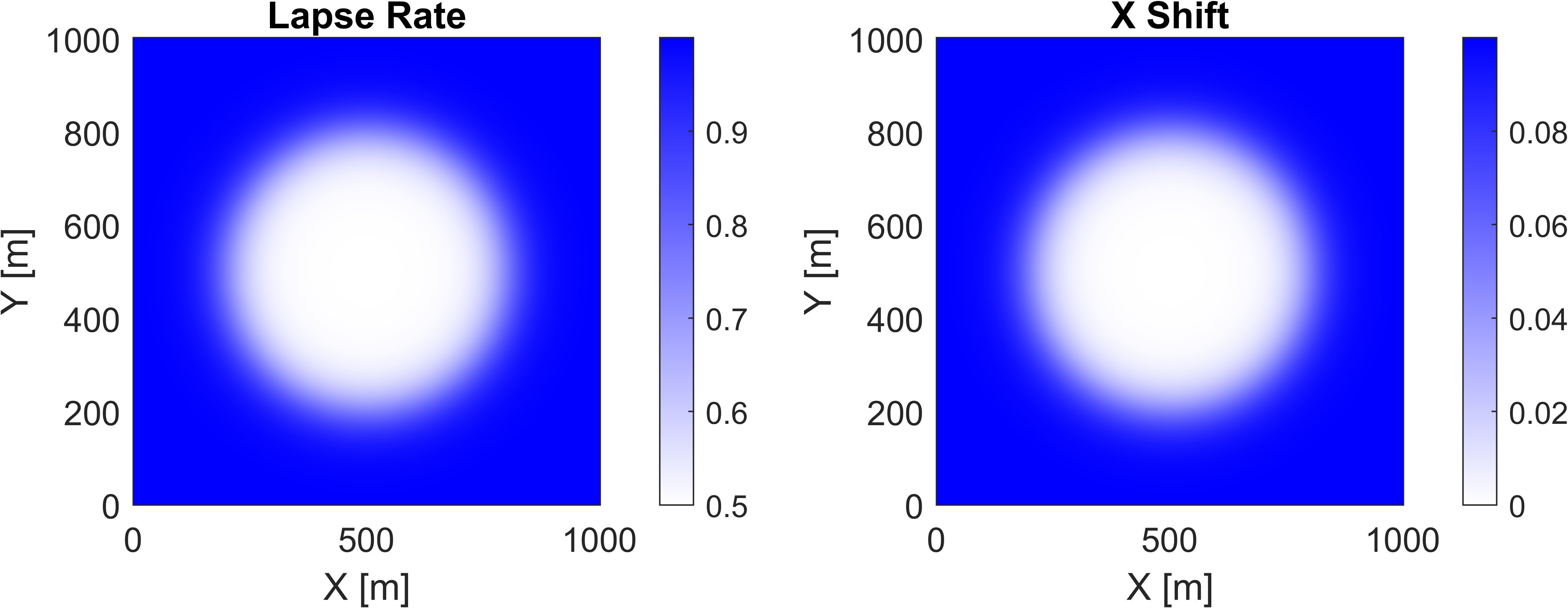}
\caption{Metric components for the comoving modified time metric. The motion of the bubble is in the +X direction. The slice is for $z=z_0$, $v_s=0.1$ c, $A_{max}=2$, $R=300$ m, $\sigma=0.015$ $\text{m}^{-1}$, $x_0=y_0=z_0=503$ m.}\label{fig:MTM_components}
\end{figure}

Bobrick and Martire explore various alterations to the Alcubierre metric in their publication titled ``Introducing Physical Warp Drives" \cite{2021CQGra..38j5009B}. Among these modifications, they specifically examine the incorporation of a dynamic lapse rate, which is detailed in section 4.5 of their work. The line element, where $\rho$ and $\theta$ are cylindrical coordinates about the axis of direction of travel $x$, are given as:

\begin{equation}
    \mathrm{d}s^2 =-\left((1-f(r_s)) \mathrm{d} t+A(r_s)^{-1} f \mathrm{~d} t\right)^2+\left(\mathrm{d} x-f(r_s) v_s \mathrm{~d} t\right)^2+\mathrm{d} \rho^2+\rho^2 \mathrm{~d} \theta^2
\end{equation}
where $A(r_s)$ is:
\begin{equation}
    A(r_s) = \begin{cases}
    A\left(r_s\right) > 1 & \text { for } \quad  r_s < R \\
    1 & \text { for } \quad r_s \geq R
    \end{cases}
\end{equation}
The inclusion of $A(r_s)$ changes the lapse rate inside the drive, hence the name ``modified time metric''. Again, this metric is analyzed in the constant velocity comoving frame. In this evaluation, $A(r_s)$ is constructed with the same fall-off function and $\sigma$ as the $f(r_s)$ function in Alcubierre. The parameters for the metric are $v_s=0.1$ c, $A_{max}=2$, $R=300$ m, $\sigma=0.015$ $\text{m}^{-1}$, $x_0=y_0=z_0=503$ m. The metric components for this warp bubble are shown in Figure \ref{fig:MTM_components}. 

\begin{figure}[hbt!]
\centering
\includegraphics[width=\textwidth]{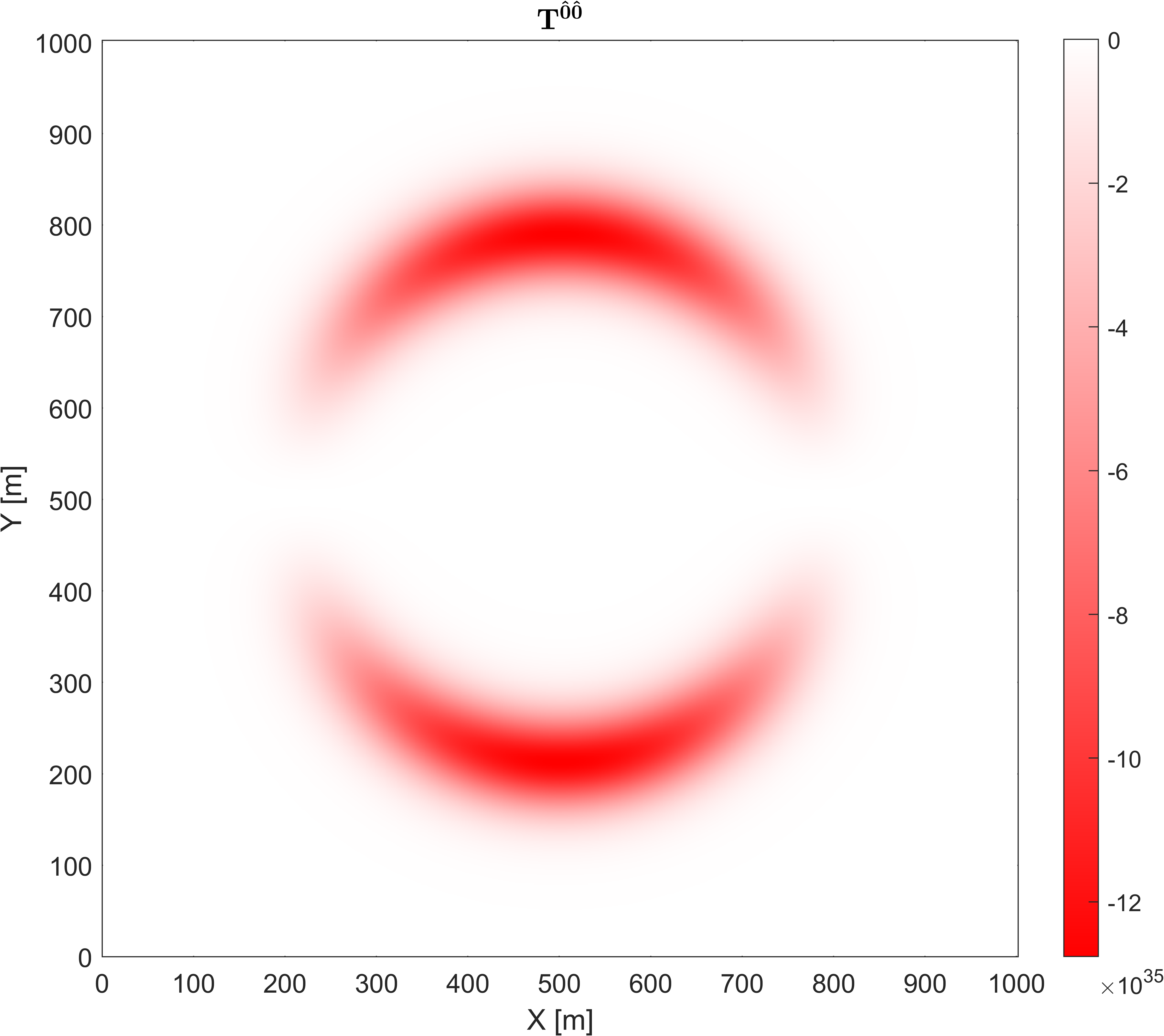}
\caption{Energy density for the modified time metric. The motion of the bubble is in the +X direction. The slice is for $z=z_0$, $v_s=0.1$ c, $A_{max}=2$, $R=300$ m, $\sigma=0.015$ $\text{m}^{-1}$, $x_0=y_0=z_0=503$ m. Units are in J/m$^3$.}\label{fig:MTM_Eden}
\end{figure}

\begin{figure}[hbt!]
\centering
\includegraphics[width=\textwidth]{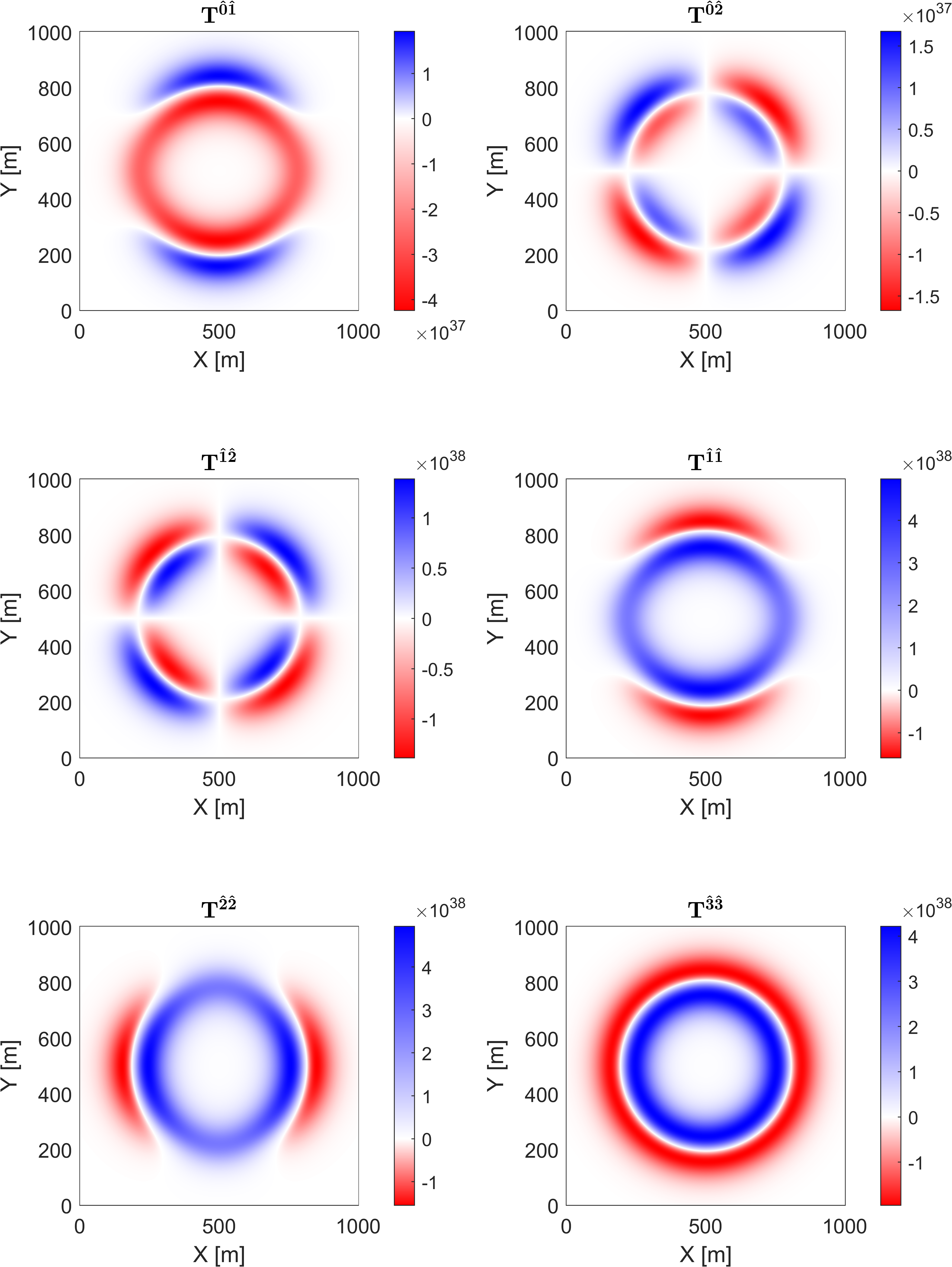}
\caption{Stress-energy tensor components for the modified time metric. The $T^{\hat{0}\hat{0}}$ component is shown in Figure \ref{fig:MTM_Eden}. The motion of the bubble is in the +X direction. The slice is for $z=z_0$, $v_s=0.1$ c, $A_{max}=2$, $R=300$ m, $\sigma=0.015$ $\text{m}^{-1}$, $x_0=y_0=z_0=503$ m. Units are in J/m$^3$.}\label{fig:MTM_Etensor}
\end{figure}

\begin{figure}[hbt!]
\centering
\includegraphics[width=0.95\textwidth]{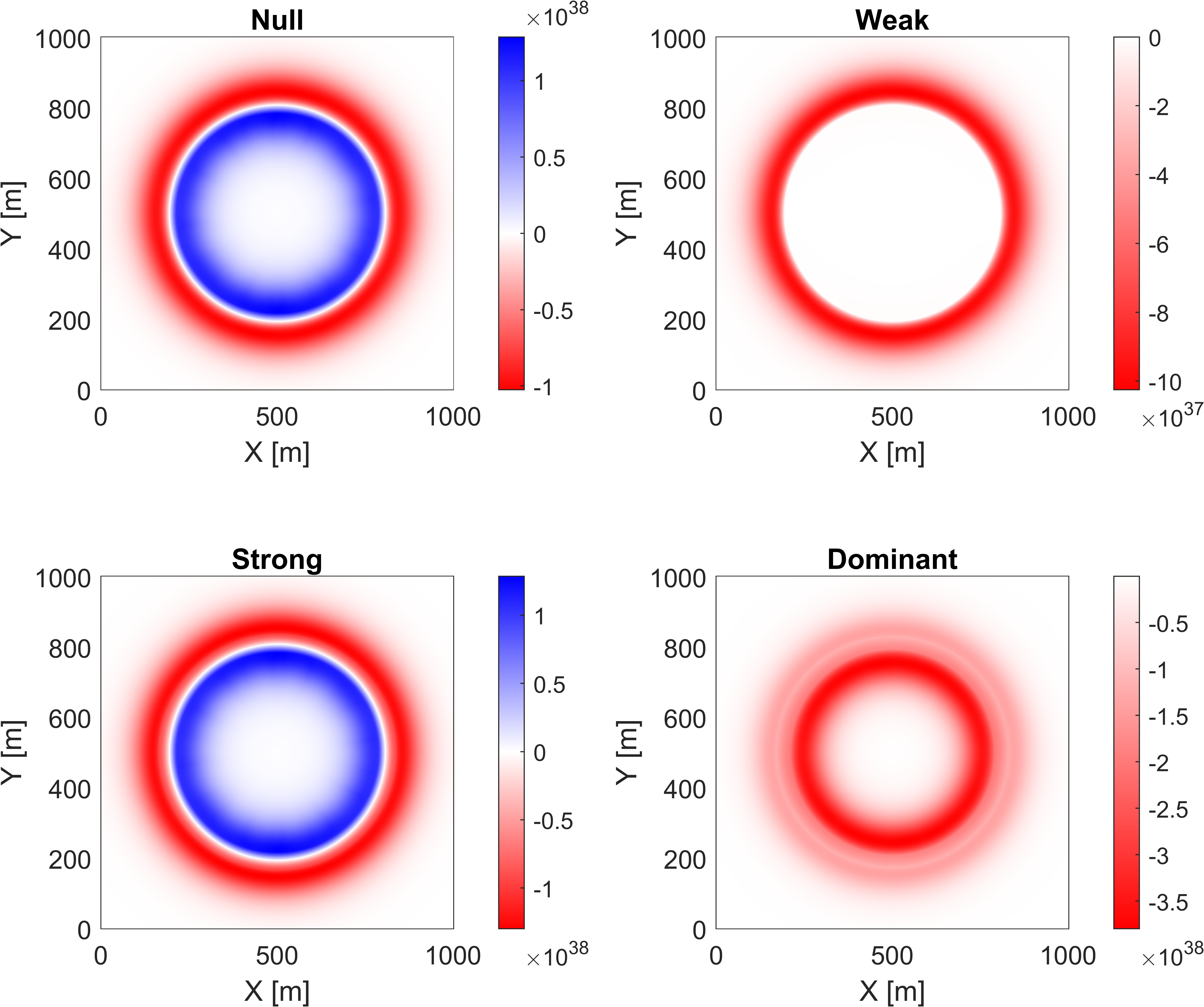}
\caption{Energy conditions for the modified-time metric. The motion of the bubble is in the +X direction. The slice is for $z=z_0$, $v_s=0.1$ c, $A_{max}=2$, $R=300$ m, $\sigma=0.015$ $\text{m}^{-1}$, $x_0=y_0=z_0=503$ m, $n_{\rm{observers}}=1000$. The minimum value among all observers is shown which can be positive. Units are in J/m$^3$.}\label{fig:MTM_Econd}
\end{figure}

\begin{figure}[hbt!]
\centering
\includegraphics[width=0.95\textwidth]{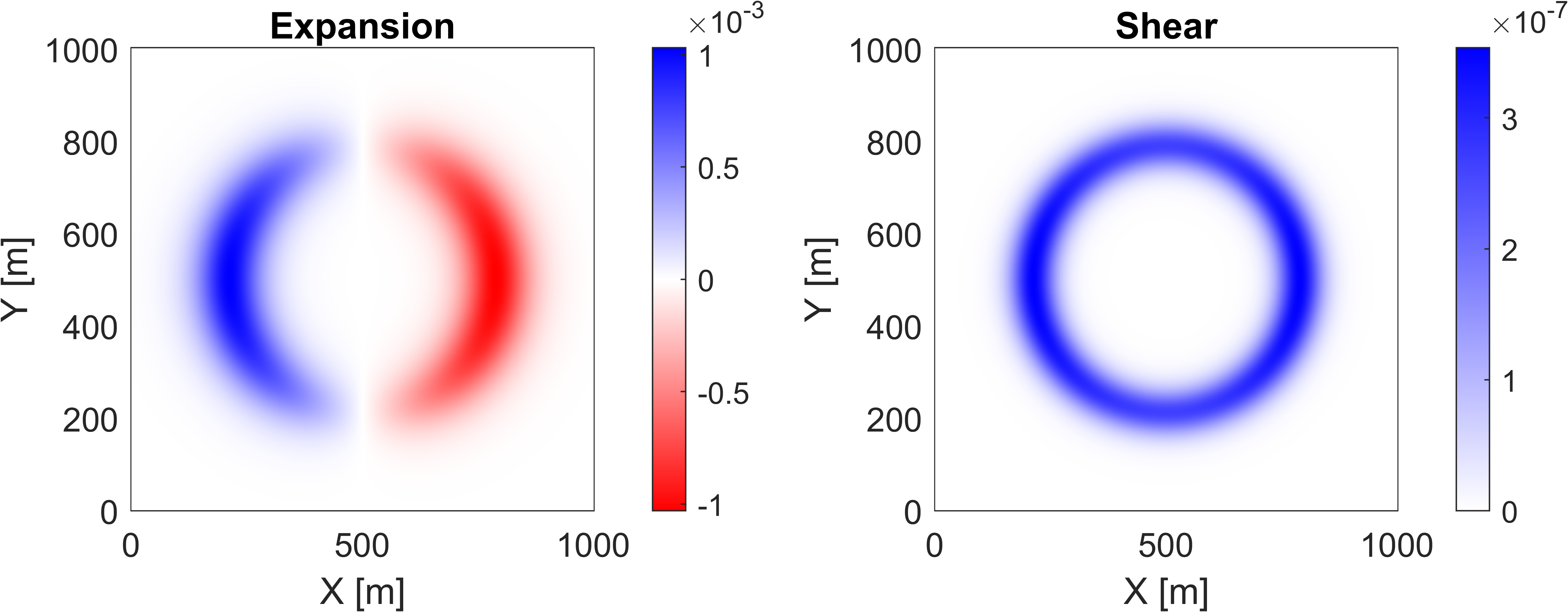}
\caption{Expansion and shear scalars for the modified time metric. The motion of the bubble is in the +X direction. The slice is for $z=z_0$, $v_s=0.1$ c, $A_{max}=2$, $R=300$ m, $\sigma=0.015$ $\text{m}^{-1}$, $x_0=y_0=z_0=503$ m.}\label{fig:MTM_scalars}
\end{figure}

Figure \ref{fig:MTM_Eden} illustrates the energy density, which exhibits similarities to the Alcubierre metric. Examining Figure \ref{fig:MTM_Etensor}, the stress-energy tensor reveals the presence of positive pressures in the interior wall of the bubble, in contrast to the negative pressures observed in the Alcubierre metric. The modification of the lapse rate results in different energy condition violations, as depicted in Figure \ref{fig:MTM_Econd}. Lastly, Figure \ref{fig:MTM_scalars} showcases the metric scalars, which exhibit similar maximum values but with broader shapes than the Alcubierre metric.


\clearpage

\subsection{Lentz-Inspired Metric}

The Lentz metric, introduced in a recent publication  \cite{2022arXiv220100652L}, presents a dramatic departure from the conventional spherically symmetric solutions proposed by Alcubierre and others. Unlike its prior warp solutions, the Lentz metric tackles head-on the challenge of avoiding violations of the WEC by incorporating shift vectors along multiple directions. Lentz's insight lies in recognizing a limitation of previous warp metrics, where a metric featuring a single shift vector component inevitably generates negative energy density for Eulerian observers. In contrast, Lentz's approach involves utilizing additional shift components derived from a ``potential" described by a linear wave equation.

\begin{figure}[ht]
\centering
\includegraphics[width=\textwidth]{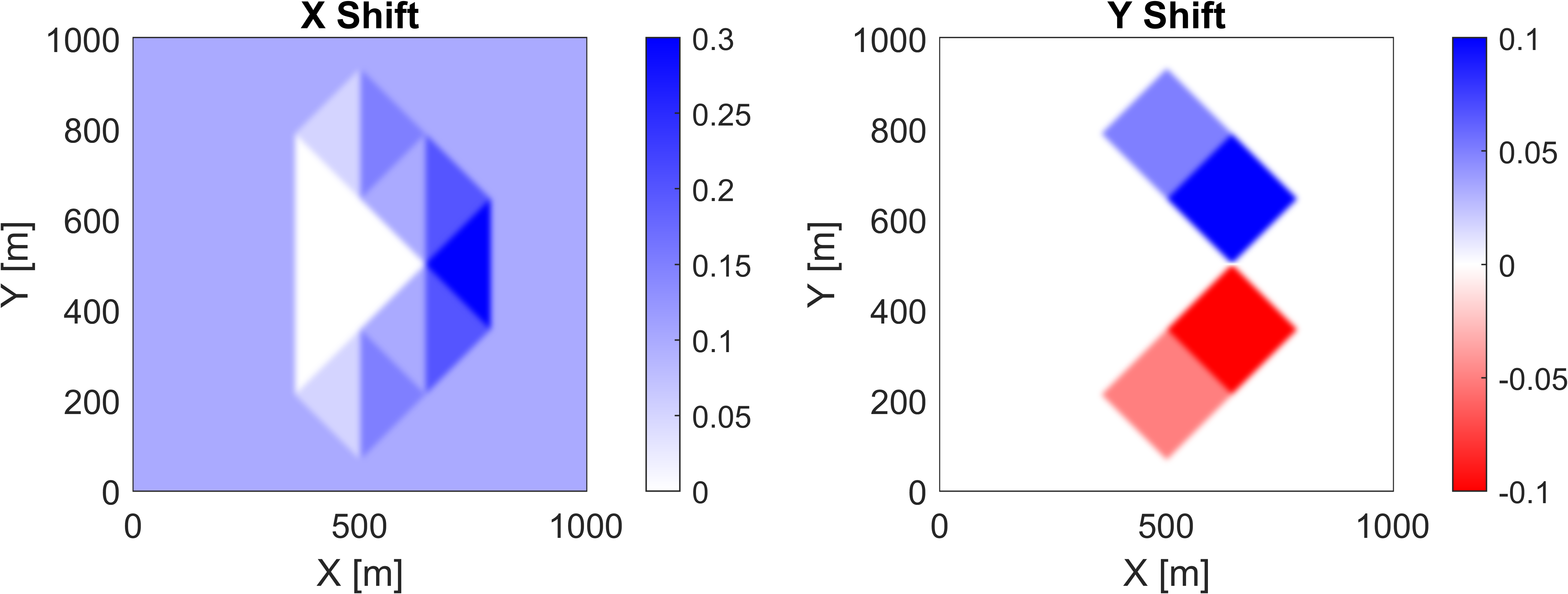}
\caption{Metric components for the comoving Lentz-inspired metric. The motion of the bubble is in the +X direction.}\label{fig:LM_components}
\end{figure}

A metric inspired by this construction was created in the constant velocity comoving frame for analysis in Warp Factory. Instead of the linear wave equation method used by Lentz, we construct a piece-wise metric using two shift vector components in the X and Y directions. The values and shape of these shift vector sections were designed to resemble Figure 1 from \cite{2022arXiv220100652L}. Once these 2D sections were created, the shift vectors were smoothed via 2D Gaussian smoothing. Figure \ref{fig:LM_components} illustrates similar shift vectors to those proposed by Lentz. The metric's parameters are $v_s=0.1$ c, initial coordinates $x_0=y_0=503$ m, and a Gaussian smoothing factor of $10$ m.

\begin{figure}
\centering
\includegraphics[width=\textwidth]{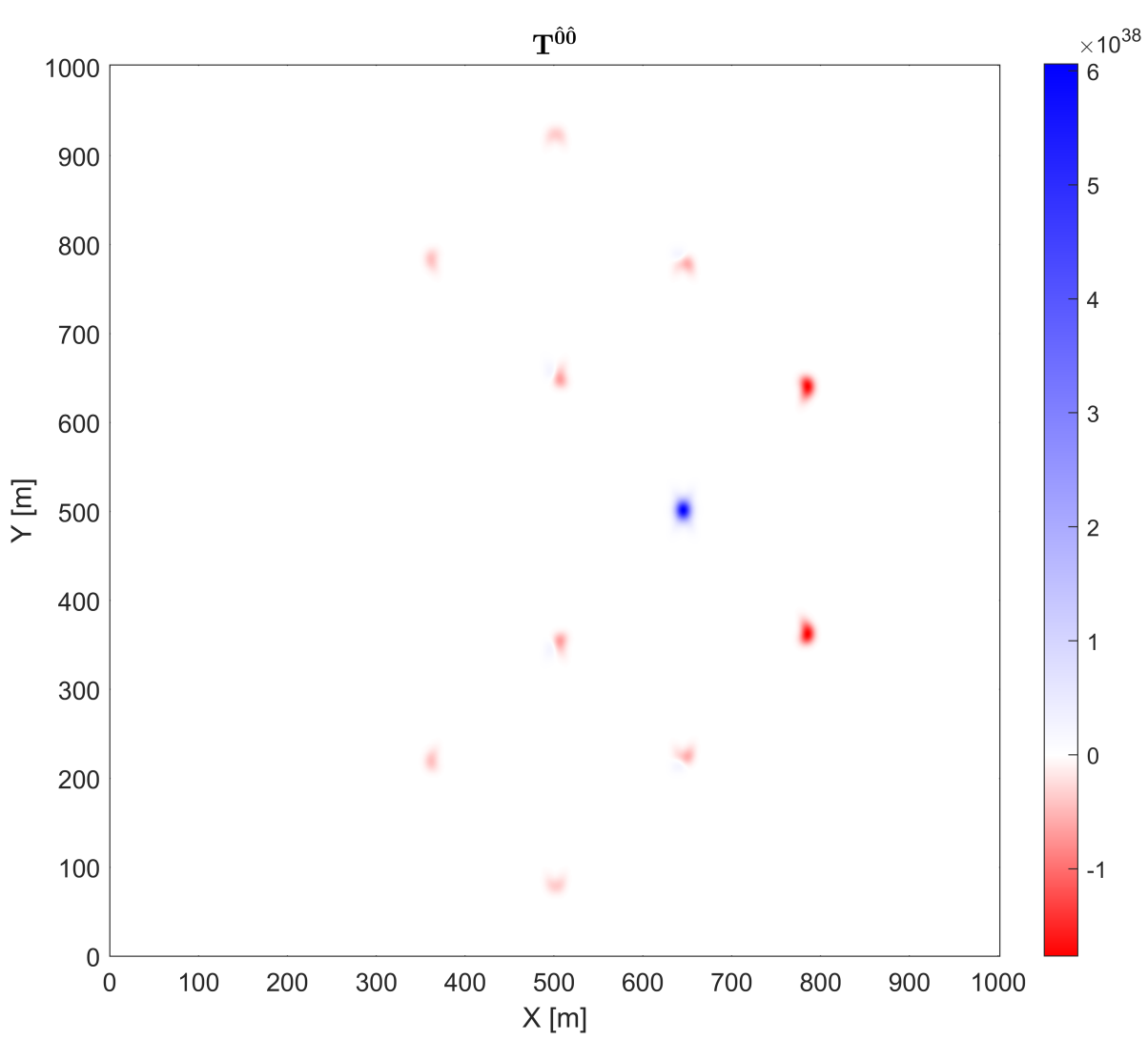}
\caption{Energy density for the Lentz-inspired metric. The motion of the bubble is in the +X direction. $v_s=0.1$ c. Units are in J/m$^3$.}\label{fig:LM_Eden}
\end{figure}

\begin{figure}
\centering
\includegraphics[width=\textwidth]{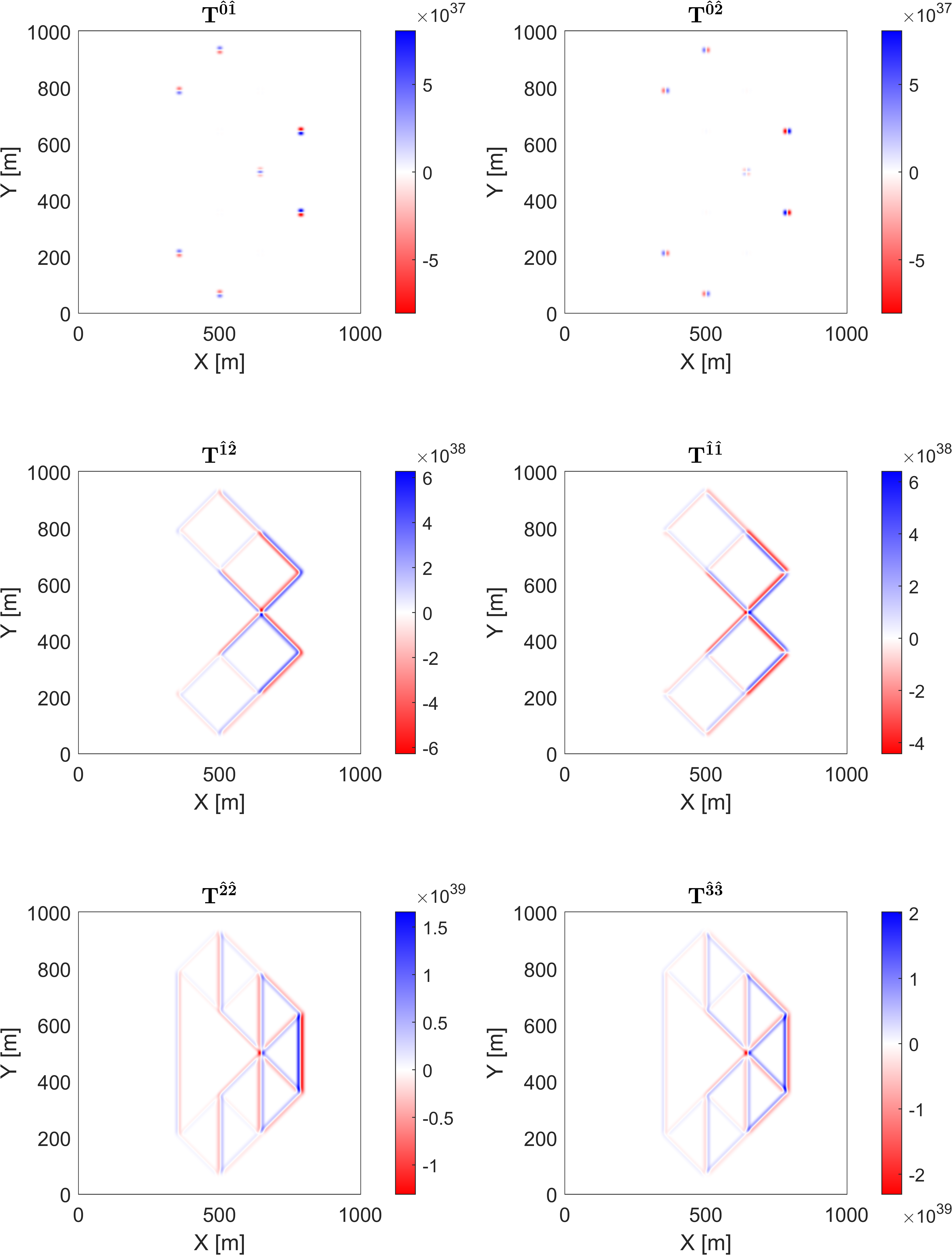}
\caption{Stress-energy tensor components for the Lentz-inspired metric. The $T^{\hat{0}\hat{0}}$ component is shown in Figure \ref{fig:LM_Eden}. The motion of the bubble is in the +X direction. $v_s=0.1$ c. Units are in J/m$^3$.}\label{fig:LM_tensor}
\end{figure}

\begin{figure}
\centering
\includegraphics[width=0.95\textwidth]{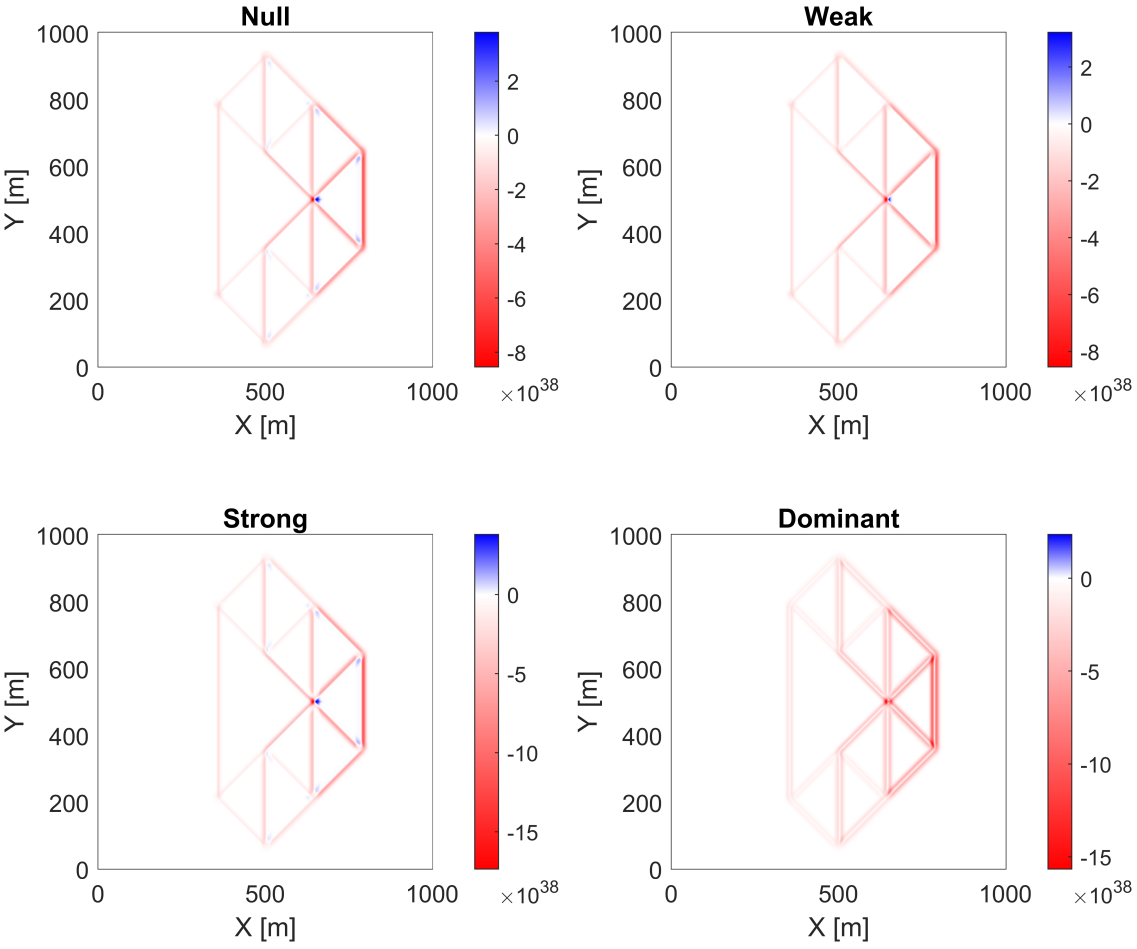}
\caption{Energy conditions for the Lentz-inspired metric. The motion of the bubble is in the +X direction. $v_s=0.1$ c and $n_{\rm{observers}}=1000$. The minimum value among all observers is shown which can be positive. Units are in J/m$^3$.}\label{fig:LM_Econd}
\end{figure}

Figure \ref{fig:LM_Eden} illustrates the evaluation of the energy density wherein the energy density remains concentrated primarily at the corners where derivatives of the metric exist. A noteworthy distinction in our metric as compared to Lentz's metric is the distribution of Eulerian energy density with ours having regions of both positive and negative energy only at the corners of the shift regions. While the energy density predominantly approaches zero, violations of the energy conditions emerge consistently at the interfaces of the rhomboid structure, as depicted in Figure \ref{fig:LM_Econd}. The cause of this violation lies in the non-zero pressure terms at all boundaries of the shift vectors, as seen in Figure \ref{fig:LM_tensor}, where the energy density is absent. This phenomenon manifests as negative energy density to observers in other timelike frames. 

\begin{figure}
\centering
\includegraphics[width=0.95\textwidth]{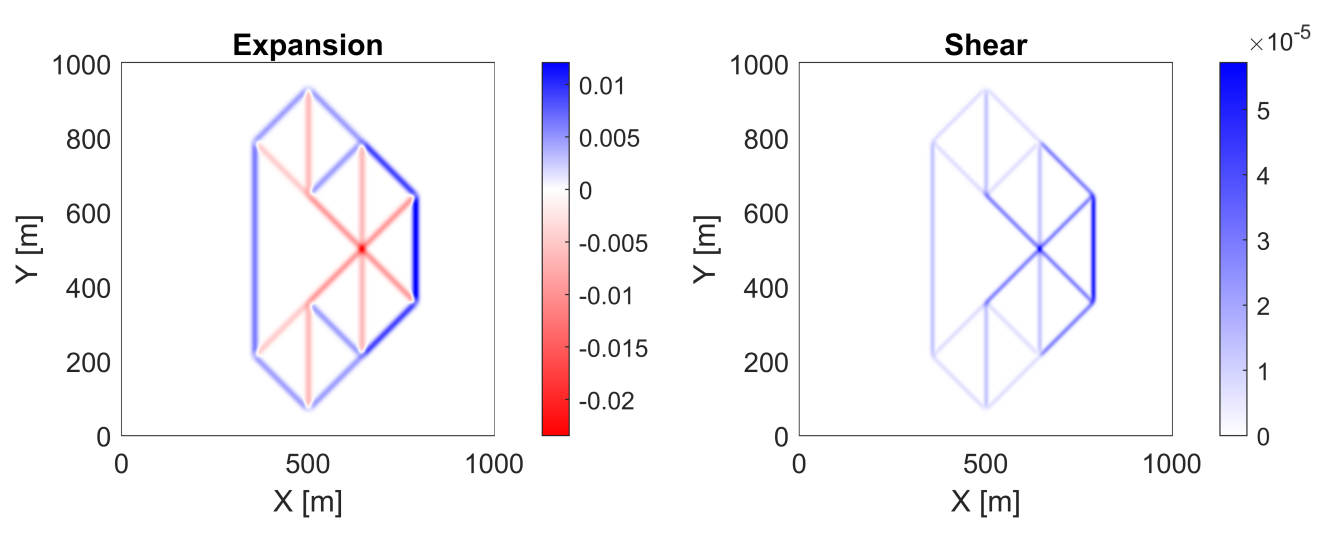}
\caption{Expansion and shear scalars for the Lentz-inspired metric. The motion of the bubble is in the +X direction. $v_s=0.1$ c.}\label{fig:LM_scalars}
\end{figure}

While the reduction of negative energy density in the Eulerian frame is important, removing violation requires consideration of the pressure, momenta, and energy density relationship, which the Lentz-inspired solution does not manage. From the work here, it is not demonstrated that no physical solution exists using this approach, but a key challenge has been highlighted for the selection of the gradient boundaries, which must generate positive energy density to overcome the non-zero pressure and momentum density to make this solution physical. Namely, any physical variant must have positive energy along all edges of the rhomboid structure. Lastly, Figure \ref{fig:LM_scalars} presents the scalars for the Lentz-inspired metric. The full construction and evaluation of the Lentz metric in Warp Factory was beyond the scope of this paper and will be left for future work.


\clearpage

\section{Discussion}\label{sec:discussion}

\subsection{Physicality Results}
For each of the metrics, we performed a full null and timelike vector field analysis of the energy conditions, with the results shown in Table \ref{tab:Econd_summary}. Generally, when considering the physicality of a warp solution, violations can be attributed to two factors: the presence of negative energy density or the occurrence of pressure and momentum density terms surpassing the energy density at specific locations in the bubble. Consequently, merely ensuring that the energy density is non-negative from the perspective of an Eulerian observer is insufficient as a criterion for discussing physicality for a metric.

\begin{table}[ht]
\centering
\caption{Comparison of proposed warp drive solution physicality as evaluated using Warp Factory. The lapse rate $\alpha$ means a constant rate whereas $\tilde{\alpha}$ is a modified lapse within the drive radius. (\textcolor{red}{\ding{55}}) indicates a violation of the corresponding energy condition at at least one point in the space.}
\vspace{0.2cm}
\label{tab:Econd_summary}
\begin{tabular}{C{5cm}C{4cm}C{1.1cm}C{1.1cm}C{1.1cm}C{1.1cm}}
    \toprule
     \ns
    Solution Name & $g_{\mu\nu}$ (pseudo-Cartesian) & NEC & WEC & DEC & SEC \\
    \ns
    \midrule 
    \textbf{Alcubierre} \cite{1994CQGra..11L..73A} & 
    $
    \begin{pmatrix}
    -\alpha^2+\beta_1^2 & \beta_1 \\
    \beta_1 & \delta_{ij} \\
    \end{pmatrix} 
    $
    & \textcolor{red}{\ding{55}} & \textcolor{red}{\ding{55}} & \textcolor{red}{\ding{55}} & \textcolor{red}{\ding{55}} \\ \noalign{\smallskip} 

    \textbf{Van Den Broeck} \cite{1999CQGra..16.3973V} & 
    $
    \begin{pmatrix} 
    -\alpha^2+\gamma^{11}\beta_1^2  & \beta_1 \\
    \beta_1 & \gamma_{ij} \\
    \end{pmatrix} 
    $
    & \textcolor{red}{\ding{55}} & \textcolor{red}{\ding{55}} & \textcolor{red}{\ding{55}} & \textcolor{red}{\ding{55}}  \\ \noalign{\smallskip} 

     \textbf{Modifed Time}  \cite{2021CQGra..38j5009B} & 
     $
    \begin{pmatrix} 
    -\tilde{\alpha}^2+\beta_1^2 & \beta_1 \\
    \beta_1 & \delta_{ij} \\
    \end{pmatrix} 
    $
    & \textcolor{red}{\ding{55}}  & \textcolor{red}{\ding{55}} & \textcolor{red}{\ding{55}} & \textcolor{red}{\ding{55}} \\ \noalign{\smallskip}

    \textbf{Lentz-Inspired} \cite{2022arXiv220100652L} & 
    $
    \begin{pmatrix} 
    -\alpha^2+\beta_{[1,2]}^2 & \beta_{[1,2]} \\
    \beta_{[1,2]} & \delta_{ij} \\
    \end{pmatrix} 
    $
    & \textcolor{red}{\ding{55}}  & \textcolor{red}{\ding{55}} & \textcolor{red}{\ding{55}} & \textcolor{red}{\ding{55}} \\ \noalign{\smallskip} 
    
    \bottomrule
\end{tabular}
\end{table}
\subsection{Scalars and Momentum Flow}

Expansion is the most frequently discussed scalar in the literature and in the general public, particularly the expansion scalar of the Alcubierre metric. However, Natario has shown that expansion is not a fundamental feature of warp drives \cite{2002CQGra..19.1157N}. While not a requirement, in the warp drives analyzed in this work, it is observed as a common characteristic that there is a contraction in front of the drive and an expansion behind it.

The metrics above all exhibit similar contraction in front and expansion behind along the direction of travel, with it occurring along the boundaries of change in the shift vector. Likewise, the shear is always positive and similarly exists along the boundaries where the shift vector changes. The Lentz-inspired metric is a more complicated case, but generally, the interior shift boundaries have a contraction and the exterior has an expansion. The Lentz-inspired metric also has shears along the boundaries of the shift vector transitions in a similar manner as the other metrics. 

With Warp Factory, visualizations of the stress-energy tensor reveal more of its structure than has been discussed before in the literature. Of particular interest are the momentum flux components. For each of the warp drives discussed here, we see a similar story of momentum structure, which has a kind of toroidal motion, especially along directions that are perpendicular to the axis of motion. This can be seen in the cross-section plots of the Alcubierre metric in which the $p_i$ stress-energy components change sign halfway through the warp bubble wall. This structure can be better visualized using the momentum flow analysis which can trace path lines of the momentum flux. Animations of the momentum flow lines for Alcubierre, Van Den Broeck, and Modified Time are provided as supplemental material to this paper.

\section{Conclusion}\label{sec:conclusion}

Warp Factory is a specialized toolkit designed for analyzing spacetimes related to warp drives. The toolkit addresses the unique analysis needs of the warp community and offers comprehensive capabilities for evaluating novel warp metrics' physicality. Traditionally, the analysis of physicality in these metrics has been limited to considering the energy density observed by Eulerian observers. However, this approach alone is insufficient for a thorough evaluation of the energy conditions. To overcome these limitations and enable the exploration of more complex solutions, Warp Factory employs a numerical framework.

By using Warp Factory, researchers can extensively explore a wide range of warp metrics, particularly in cases where analytical calculations become impractical. The capabilities of Warp Factory are demonstrated in this paper by analyzing a handful of popular metrics. Using its numerical approach, Warp Factory provides researchers with new insights including 3D visualizations and its ability to do comprehensive analysis of stress-energy tensors. A notable finding in this work is a demonstration that metrics like the Lentz-inspired solution presented here, which may exhibit solely positive Eulerian energy density, can still violate the weak energy condition when examined across all timelike observers. 

Using Warp Factory the APL team has also been exploring more warp spacetimes. In a separate paper, currently in preparation, a new constant velocity warp drive solution has been found without energy condition violations \cite{PhysicalWarp2023}. The discovery of this exciting result was made possible by utilizing the tools provided by Warp Factory, as described in this paper.\\

\noindent Warp Factory can be found at \url{https://github.com/NerdsWithAttitudes/WarpFactory}.


\clearpage

\appendix

\clearpage

\section{Solving for the Tetrad Components}\label{apx:euleriantransformation}
Assume that $e_{\hat{\mu}}^\mu$ is a lower triangular matrix (motivated in Section \ref{sec:stressenergy}).

\begin{equation}
    e^\mu_{\hat{\mu}} =
    \begin{pmatrix}
    \vrule & \vrule & \vrule & \vrule  \\ 
    e^{\mu}_0 & e^{\mu}_1 & e^{\mu}_2   & e^{\mu}_3 \\
    \vrule & \vrule & \vrule & \vrule 
    \end{pmatrix} = 
    \begin{pmatrix}
        e_1 & 0 & 0 & 0 \\
        e_2 & e_5 & 0 & 0 \\
        e_3 & e_6 & e_8 & 0 \\
        e_4 & e_7 & e_9 & e_{10} \\
    \end{pmatrix}
\end{equation}
The metric $g_{\mu\nu}$ is given by its components:
\begin{equation}
    g_{\mu\nu} =
    \begin{pmatrix}
        g_1 & g_2 & g_3 & g_4 \\
        g_2 & g_5 & g_6 & g_7 \\
        g_3 & g_6 & g_8 & g_9 \\
        g_4 & g_7 & g_9 & g_{10} \\
    \end{pmatrix}
\end{equation}
Rearranging the tetrad equation:
\begin{equation}
    g^{\mu\nu} = e^\mu_{\hat{\mu}} e^\nu_{\hat{\nu}} \eta^{\hat{\mu}\hat{\nu}} \implies  e^\mu_{\hat{\mu}} e^\nu_{\hat{\nu}} g_{\mu\nu} = \eta_{\hat{\mu} \hat{\nu}}
\end{equation}
Or, in matrix notation,
\begin{equation}
    \textbf{E}^T \ \textbf{g} \ \textbf{E} = \boldsymbol{\eta}
\end{equation}
Solving this system of equations for $e^\mu_{\hat{\mu}}$ in terms of $g_{\mu\nu}$ gives us:

\begin{equation}
\begin {split}
    e_0^\mu &= 
    \sqrt{\frac{1}{C(-D)}}
    \begin{pmatrix}
    -C\\
    g_{2}\,{g_{9}}^2+g_{3}\,g_{6}\,g_{10}+g_{4}\,g_{7}\,g_{8}-g_{2}\,g_{8}\,g_{10}-g_{3}\,g_{7}\,g_{9}-g_{4}\,g_{6}\,g_{9}\\
    g_{3}\,{g_{7}}^2+g_{2}\,g_{6}\,g_{10}+g_{4}\,g_{5}\,g_{9}-g_{2}\,g_{7}\,g_{9}-g_{3}\,g_{5}\,g_{10}-g_{4}\,g_{6}\,g_{7}\\
    g_{4}\,{g_{6}}^2+g_{2}\,g_{7}\,g_{8}+g_{3}\,g_{5}\,g_{9}-g_{2}\,g_{6}\,g_{9}-g_{3}\,g_{6}\,g_{7}-g_{4}\,g_{5}\,g_{8}
    \end{pmatrix}\\
    e_1^\mu &= 
    \sqrt{\frac{1}{BC}}
    \begin{pmatrix}
    0\\
    B\\
    g_{6}\,g_{10}-g_{7}\,g_{9}\\
    g_{6}\,g_{9}-g_{7}\,g_{8}
    \end{pmatrix}\\
    e_2^\mu &= 
    \sqrt{\frac{1}{AB}}
    \begin{pmatrix}
    0\\
    0\\
    A\\
    -g_9
    \end{pmatrix}\\
    e_3^\mu &= \sqrt{\frac{1}{A}}
    \begin{pmatrix}
    0\\
    0\\
    0\\
    1
    \end{pmatrix}
        \end{split}
\end{equation}

\clearpage

where $A$, $B$, $C$, and $D$ are the \textit{leading principle minors} of $g_{\mu\nu}$ and are given by:
\begin{equation}
\begin{split}
    A =&\  g_{10}\\
    B =& \left(-{g_{9}}^2+g_{8}\,g_{10}\right)\\
    C =& \left(-g_{5}\,{g_{9}}^2-g_{8}\,{g_{7}}^2-g_{10}\,{g_{6}}^2+2\,g_{6}\,g_{7}\,g_{9}+g_{5}\,g_{8}\,g_{10}\right)\\
    D =& \left({g_{2}}^2\,{g_{9}}^2+{g_{3}}^2\,{g_{7}}^2+{g_{4}}^2\,{g_{6}}^2\right. \\
    & -g_{8}\,g_{10}\,{g_{2}}^2-g_{5}\,g_{10}\,{g_{3}}^2-g_{5}\,g_{8}\,{g_{4}}^2-g_{1}\,g_{10}\,{g_{6}}^2-g_{1}\,g_{8}\,{g_{7}}^2-g_{1}\,g_{5}\,{g_{9}}^2 \\
    & +2\,g_{10}\,g_{2}\,g_{3}\,g_{6}+2\,g_{8}\,g_{2}\,g_{4}\,g_{7}+2\,g_{5}\,g_{3}\,g_{4}\,g_{9}+2\,g_{1}\,g_{6}\,g_{7}\,g_{9} \\
    & -2\,g_{2}\,g_{3}\,g_{7}\,g_{9}-2\,g_{2}\,g_{4}\,g_{6}\,g_{9}-2\,g_{3}\,g_{4}\,g_{6}\,g_{7} \\
    & \left.+g_{1}\,g_{5}\,g_{8}\,g_{10}\right)
\end{split}
\end{equation}

In order to prevent imaginary values in the transformations, the metric signature, and therefore the values of (D,C,B,A), must be (-,+,+,+).

\section{Error Discussion}\label{apx:errors}

\begin{figure}[ht]
\centering
\includegraphics[width=0.9\textwidth]{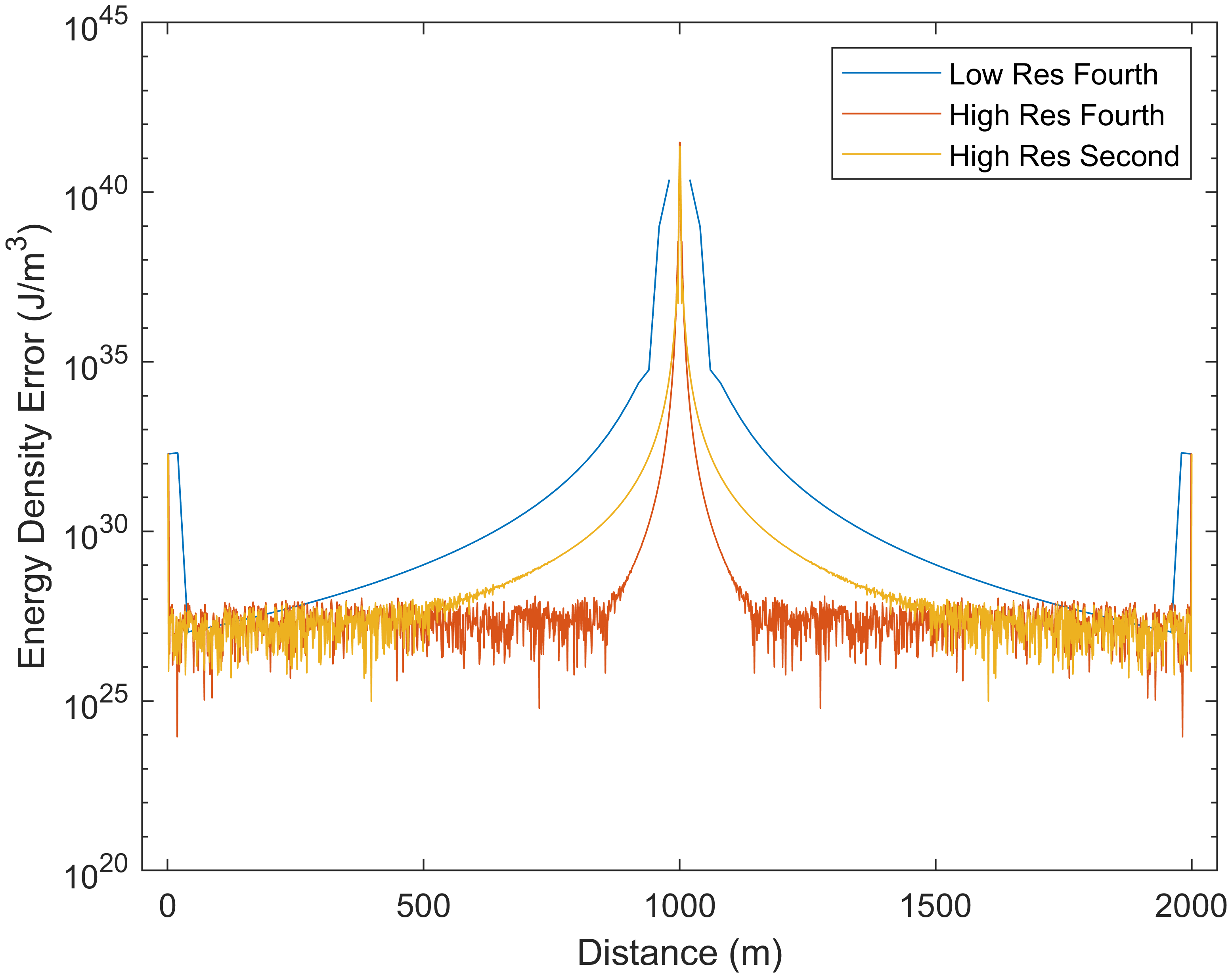}
\caption{Plots of errors of Warp Factory's numerical evaluations for a Schwarzschild solution. The error is computed by comparing the numerical $T^{00}$ value to its true value which is zero. The low-resolution grid is 20 m, and the high-resolution grid is 1 m. The center of the Schwarzschild metric is at 1000 meters and its radius is 0.01 meters.}\label{fig:errors}
\end{figure}

All 4 types of errors can be seen in Figure \ref{fig:errors}.
The Schwarzschild metric acts well to highlight the different types of numerical error of the code since, outside of the Schwarzschild radius, the evaluated energy density should be 0. Therefore, any computed non-zero energy density is due to some type of error.\\

\subsection{Edge of Grid Error}
The first, and perhaps most important error from a practical standpoint is edge of grid error. The energy tensor is returned in the same grid size as the input metric. At the edges of the world, a full finite difference cannot be taken as the points needed to evaluate the derivatives at the edges are outside of the world. This can be seen in the edges of the plot where all three evaluations have considerable energy density spikes at the very bounds of the space. For energy tensor and energy condition plotting, it is recommended to exclude a boundary of 2 grid points at the edges of the space.

\subsection{Finite Difference Discretization Error}
This type of error occurs when the rate of change of the metric components nears or exceeds the grid resolution. The error is a direct function of grid resolution. As can be seen in the plot, the blue low-resolution line has more error over a wider area than the higher-resolution orange line. Even with high resolution, though, the orange line does exhibit this type of error as it nears the center of the Schwarzschild metric. Metrics with sharp transitions can lead to these finite difference edge effects and may give energy/violation that wouldn't actually exist at those boundaries.

\subsection{Floating Point Round Off Error}
For double-value precision, there is a maximum dynamic range of numbers that can be stored. As can be seen in Figure \ref{fig:errors}, the errors become very noisy around 10$^{28}$. The double precision limit prohibits storing more discrete values. This precision limit is based on the maximum value that exists in the array that is storing the values.

\subsection{Finite Difference Truncation Error}
This error results from the finite difference not taking a perfect derivative of the function. The order of the finite difference (second, fourth, etc.) dictates how accurately the real derivative is evaluated. As can be seen in Figure \ref{fig:errors}, the yellow second-order finite difference has higher errors than the orange fourth-order finite difference.

\clearpage
\subsection{Alcubierre Numerical Errors}
\begin{figure}[ht]
\centering
\includegraphics[width=1\textwidth]{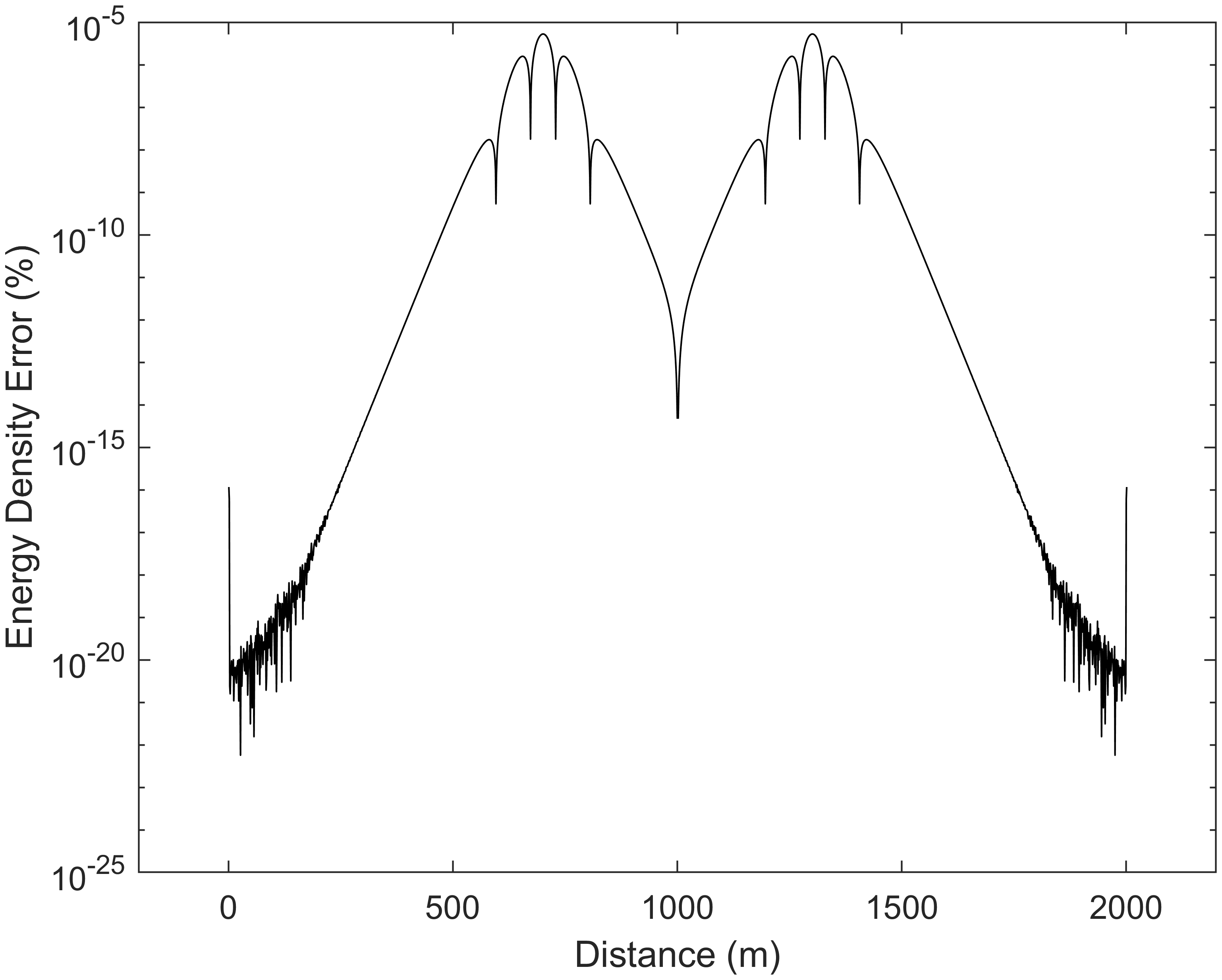}
\caption{Percent error for the numerical evaluation with a 1-meter grid resolution of the Alcubierre metric compared to an exact analytical evaluation. The numerical error is given as a percent of the maximum energy density of the analytical evaluation.}\label{fig:alc_errors}
\end{figure}

Figure \ref{fig:alc_errors} shows the energy density errors versus the exact analytical expression for the energy density of an Alcubierre warp bubble. The radius R, velocity v, and $\sigma$ are the same as in Figure \ref{fig:AM_components}. The bubble here is centered at 1000 meters and the space is extended to show how the errors fall off.

\clearpage
\section*{References}
\bibliography{papermain.bib}{}

\begin{thebibliography}{10}

\bibitem{1994CQGra..11L..73A}
Miguel {Alcubierre}.
\newblock {LETTER TO THE EDITOR: The warp drive: hyper-fast travel within
  general relativity}.
\newblock {\em Classical and Quantum Gravity}, 11(5):L73--L77, May 1994.

\bibitem{2021CQGra..38j5009B}
Alexey {Bobrick} and Gianni {Martire}.
\newblock {Introducing physical warp drives}.
\newblock {\em Classical and Quantum Gravity}, 38(10):105009, May 2021.

\bibitem{2004sgig.book.....C}
Sean~M. {Carroll}.
\newblock {\em {Spacetime and geometry. An introduction to general
  relativity}}.
\newblock 2004.

\bibitem{1997PhRvD..56.2100E}
Allen~E. {Everett} and Thomas~A. {Roman}.
\newblock {Superluminal subway: The Krasnikov tube}.
\newblock {\em Physical Review D}, 56(4):2100--2108, August 1997.

\bibitem{2021CQGra..38o5020F}
Shaun D.~B. {Fell} and Lavinia {Heisenberg}.
\newblock {Positive energy warp drive from hidden geometric structures}.
\newblock {\em Classical and Quantum Gravity}, 38(15):155020, August 2021.

\bibitem{1995PhRvD..51.4277F}
L.~H. {Ford} and Thomas~A. {Roman}.
\newblock {Averaged energy conditions and quantum inequalities}.
\newblock {\em Physical Review D}, 51(8):4277--4286, April 1995.

\bibitem{PhysicalWarp2023}
Jared. {Fuchs}, Christopher. {Helmerich}, Alexey. {Bobrick}, Luke. {Sellers},
  Brandon. {Melcher}, and Gianni. {Martire}.
\newblock {Constant Velocity Physical Warp Drive Solution}.
\newblock {\em Manuscript in preparation}, 2023.

\bibitem{2007gr.qc.....3035G}
Eric {Gourgoulhon}.
\newblock {3+1 Formalism and Bases of Numerical Relativity}.
\newblock {\em arXiv e-prints}, pages gr--qc/0703035, March 2007.

\bibitem{hawking_ellis_1973}
S.~W. Hawking and G.~F.~R. Ellis.
\newblock {\em The Large Scale Structure of Space-Time}.
\newblock Cambridge Monographs on Mathematical Physics. Cambridge University
  Press, 1973.

\bibitem{2020CQGra..37s3001K}
Eleni-Alexandra {Kontou} and Ko~{Sanders}.
\newblock {Energy conditions in general relativity and quantum field theory}.
\newblock {\em Classical and Quantum Gravity}, 37(19):193001, October 2020.

\bibitem{2020arXiv200607125L}
Erik~W. {Lentz}.
\newblock {Breaking the Warp Barrier: Hyper-Fast Solitons in
  Einstein-Maxwell-Plasma Theory}.
\newblock {\em arXiv e-prints}, page arXiv:2006.07125, June 2020.

\bibitem{2022arXiv220100652L}
Erik~W. {Lentz}.
\newblock {Hyper-Fast Positive Energy Warp Drives}.
\newblock {\em arXiv e-prints}, page arXiv:2201.00652, December 2021.

\bibitem{2002CQGra..19.1157N}
Jos{\'e} {Nat{\'a}rio}.
\newblock {Warp drive with zero expansion}.
\newblock {\em Classical and Quantum Gravity}, 19(6):1157--1165, March 2002.

\bibitem{2022PhRvD.105f4038S}
Jessica {Santiago}, Sebastian {Schuster}, and Matt {Visser}.
\newblock {Generic warp drives violate the null energy condition}.
\newblock {\em Physical Review D}, 105(6):064038, March 2022.

\bibitem{shibata_2016}
Masaru Shibata.
\newblock {\em Numerical relativity}.
\newblock World Scientific Publishing Co. Pte. Ltd., 2016.

\bibitem{1999CQGra..16.3973V}
Chris {Van Den Broeck}.
\newblock {A `warp drive' with more reasonable total energy requirements}.
\newblock {\em Classical and Quantum Gravity}, 16(12):3973--3979, December
  1999.

\end{thebibliography}
\bibliographystyle{plain}

\end{document}